\documentclass[twocolumn,english,showpacs,preprintnumbers,amsmath,amssymb,floatfix,nofootinbib]{revtex4-1}

\usepackage[T1]{fontenc}
\usepackage[latin9]{inputenc}
\usepackage{color}
\usepackage{array}
\usepackage{amstext}
\usepackage{graphicx}
\usepackage{esint}
\usepackage{rotating}
\usepackage{slashed}
\usepackage{siunitx}
\usepackage[normalem]{ulem}
\usepackage[font=small,labelfont=bf]{caption}
\usepackage{xcolor}
\usepackage[colorlinks = true,
linkcolor = magenta,
urlcolor  = blue,
citecolor = red,
anchorcolor = blue]{hyperref}

\newcommand{\pom} {I\!\!P}
\newcommand{\reg} {I\!\!R}
\newcommand{\xpom}{x_{\xpom}}

\makeatletter

\providecommand{\tabularnewline}{\\}

\@ifundefined{textcolor}{}
{%
 \definecolor{BLACK}{gray}{0}
 \definecolor{WHITE}{gray}{1}
 \definecolor{RED}{rgb}{1,0,0}
 \definecolor{GREEN}{rgb}{0,1,0}
 \definecolor{BLUE}{rgb}{0,0,1}
 \definecolor{CYAN}{cmyk}{1,0,0,0}
\definecolor{MAGENTA}{cmyk}{0,1,0,0}
 \definecolor{YELLOW}{cmyk}{0,0,1,0}
 }

\@ifundefined{definecolor}
{\usepackage{color}}{}
\@ifundefined{definecolor}
{\usepackage{color}}{}
\makeatother
\usepackage{babel}

\def\Re{{\cal R \mskip-4mu \lower.1ex \hbox{\it e}\,}}
\def\Im{{\cal I \mskip-5mu \lower.1ex \hbox{\it m}\,}}

\def\tev{\,{\ifmmode\mathrm {TeV}\else TeV\fi}}
\def\gev{\,{\ifmmode\mathrm {GeV}\else GeV\fi}}
\def\mev{\,{\ifmmode\mathrm {MeV}\else MeV\fi}}
\def\to{\rightarrow}

\begin{document}


\title { Role of higher twist effects in diffractive DIS and determination of diffractive parton distribution functions }

\author{ Atefeh Maktoubian$^{1}$ }
\email{ Atefeh.Maktoubian@semnan.ac.ir }

\author{ Hossein Mehraban$^{1}$}
\email{ Hmehraban@semnan.ac.ir}

\author{ Hamzeh Khanpour$^{2,3}$ }
\email{ Hamzeh.Khanpour@cern.ch }

\author{ Muhammad Goharipour$^{3}$ }
\email{ Muhammad.Goharipour@ipm.ir }

\affiliation {
$^{(1)}$Faculty of Physics, Semnan University, P.O.Box 35131-19111, Semnan, Iran  \\
$^{(2)}$Department of Physics, University of Science and Technology of Mazandaran, P.O.Box 48518-78195, Behshahr, Iran    \\ 
$^{(3)}$School of Particles and Accelerators, Institute for Research in Fundamental Sciences (IPM), P.O.Box 19395-5531, Tehran, Iran 
}

\date{\today}

%
\begin{abstract}\label{abstract}

The current analysis aims to present the results of a QCD analysis of diffractive parton distribution functions (diffractive PDFs) at next-to-leading order (NLO) accuracy in perturbative QCD. In this new determination of diffractive PDFs, we use all available and up-to-date diffractive deep inelastic scattering (diffractive DIS) datasets from H1 and ZEUS Collaborations at HERA including the most recent H1/ZEUS combined measurements.
In this analysis, we consider the heavy quark contributions to the diffractive DIS in the so-called framework of {\tt FONLL} general mass variable flavor number scheme (GM-VFNS). 
The uncertainties on the diffractive PDFs are calculated using the standard ``Hessian error propagation'' which served to provide a more realistic estimate of the uncertainties. This analysis are enriched, for the first time, by including the nonperturbative higher twist (HT) effects in the calculation of diffractive DIS cross sections which are particularly important at large-$x$ and low $Q^{2}$ regions. Then, the stability and reliability of the extracted diffractive PDFs are investigated upon inclusion of HT effects. We discuss the novel aspects of the approach used in this QCD fit, namely, optimized and flexible parameterizations of diffractive PDFs, the inclusion of HT effects, and considering the recent H1/ZEUS combined dataset. Finally, we present the extracted diffractive PDFs with and without the presence of HT effects, and discuss the fit quality and the stability upon variations of the kinematic cuts and the fitted datasets. We show that the inclusion of HT effects in diffractive DIS can improve the description of the data which leads, in general, to a very good agreement between data and theory predictions.

\end{abstract}
%


\maketitle

\tableofcontents{}

%
\section{ Introduction and Motivation }\label{sec:one}
%

Over the past few decades, it has been known that the parton distribution functions (PDFs) of nucleons are an essential ingredient for the interpretation and quantum chromodynamics (QCD) phenomenology of hadron structure in high energy experiments such as the deep-inelastic lepton-nucleon scattering ($\ell p$ DIS) and hadron-hadron collisions. Despite the active experimental and theoretical investigations, the determination of PDFs along with their uncertainties through a global QCD analysis is still an important topic in high energy physics. In recent years, there has been an increasing amount of literature on this topic. For more detailed discussions, we refer the reader to Refs.~\cite{Gao:2017yyd,AbdulKhalek:2019bux,Khalek:2018mdn,Rottoli:2018nma,Lin:2017snn,Schmookler:2019nvf} for introductory texts on the fundamentals of QCD factorization, global QCD PDFs analyses, and phenomenological applications of PDFs in the LHC era.

Among the high energy experiments, the most interesting physics results mainly came from the H1 and ZEUS experiments at HERA-I and HERA-II which have provided an impressive wealth of information on the proton structure. In addition, among the high energy processes of interest, diffractive DIS which contributes a fraction of order 8\%-10\% to the total DIS cross section also aims to discover the underlying structure of hadrons through diffractive processes~\cite{Rasmussen:2018dgo,Helenius:2019gbd,Britzger:2018zvv}. The method of global QCD analysis for diffractive PDFs is the same as the ordinary PDFs. It is also based on QCD factorization of physical observables. According to the factorization theorem for diffractive
DIS in perturbative QCD~\cite{Collins:1996fb,Collins:2001ga,Collins:1997sr}, the hard scattering cross sections can be expressed as a convolution
between the hard partonic cross sections, which can be calculated in perturbative QCD, and the nonperturbative diffractive PDFs. The later need to be extracted from QCD analyses of variety of available hard scattering diffractive DIS experimental datasets, though the inclusion of other diffractive data from collider experiments can improve their uncertainties to a significant extent.

Recent progresses in global QCD analysis of diffractive PDFs used all available high-precision measurements from H1 and ZEUS Collaborations at HERA which have led to a precise determination of diffractive PDFs. In recent years, a considerable amount of literature has been published on diffractive PDFs analysis with their uncertainties. These include the most recent analysis by {\tt GKG18-DPDF}~\cite{Goharipour:2018yov} which has been done in the framework of {\tt xFitter}~\cite{Alekhin:2014irh} considering the most recent H1/ZEUS combined dataset~\cite{Aaron:2012hua}, {\tt H1-2006-DPDF}~\cite{Aktas:2006hy} and {\tt ZEUS-2010-DPDF}~\cite{Chekanov:2009aa}. All these analyses were performed at NLO accuracy in QCD. More recently, {\tt HK19-DDPF}~\cite{Khanpour:2019pzq} has reported sets of diffractive PDFs at NLO and, for the first time, at next-to-next-to-leading order (NNLO) accuracy in perturbative QCD by analyzing all available and up-to-date datasets for diffractive DIS including the H1/ZEUS combined measurements on hard scattering diffractive cross sections~\cite{Aaron:2012hua}. 

The current analysis aims to clarify whether the inclusion of nonperturbative higher twist (HT) effects can affect the QCD analysis of diffractive DIS data and hence can fill a gap in the QCD analysis of diffractive PDFs in literature. To this end, we analyzed the diffractive DIS datasets through QCD analyses with or without the inclusion of HT terms.
We observed that: (1) the HT effects may have significant contributions in diffractive DIS cross sections, (2) considering HT effects in the QCD analysis of diffractive PDFs could affect the data/theory comparisons and significantly improve the fit quality, and (3) could also affect the shape and size of the extracted diffractive PDFs. From the comparisons presented in this study, we see a number of interesting similarities and differences between these two diffractive PDF sets and their uncertainties. We will return to this issue in more details in Section.~\ref{sec:Results}.

The outline of this article is the following.
In Section.~\ref{sec:Phenomenological-framework} we introduce the theoretical framework and kinematical variables used for the definition of diffractive DIS processes, diffractive structure functions and diffractive reduced cross sections. In this section we also present our parameterization for the quark and gluon diffractive PDFs, the heavy quark contributions in the GM-VFNS~\cite{Harland-Lang:2014zoa,Thorne:1997ga,Thorne:2006qt} and finally the formalism of HT effects used in this analysis.  Section.~\ref{sec:Difrractive-DIS-data-sets} includes a comprehensive introduction of the diffractive DIS experimental datasets that are used in our diffractive PDFs analysis. 
The main results and findings of the present diffractive PDFs analysis are discussed in details in Section.~\ref{sec:Results}. In the same section, we present the extracted diffractive PDFs and detailed comparisons with other results in literature. Then, we assess the quality of our fit by comparing the resulting diffractive reduced cross sections with the experimental datasets. We also assess the stability of our QCD analysis with respect to the inclusion of nonperturbative HT terms.
Lastly, in Section.~\ref{sec:Discussion}, we summarize and discuss our analysis. This section also includes a brief discussion of the implication of the findings to future research.

%
\section{ Phenomenological framework }\label{sec:Phenomenological-framework}
%

In this section we discuss in details the theoretical framework for the evaluation of diffractive DIS structure functions and reduced cross sections, the quark and gluon diffractive PDFs at the input scale, the heavy quark mass effects, the nonperturbative HT corrections, and finally the software tools that used for the numerical calculations of diffractive DIS cross sections.

%
\subsection{ Theory Settings }\label{sec:Theory-Settings}
%

\begin{figure}[t!]
\vspace{0.20cm}
\resizebox{0.450\textwidth}{!}{\includegraphics{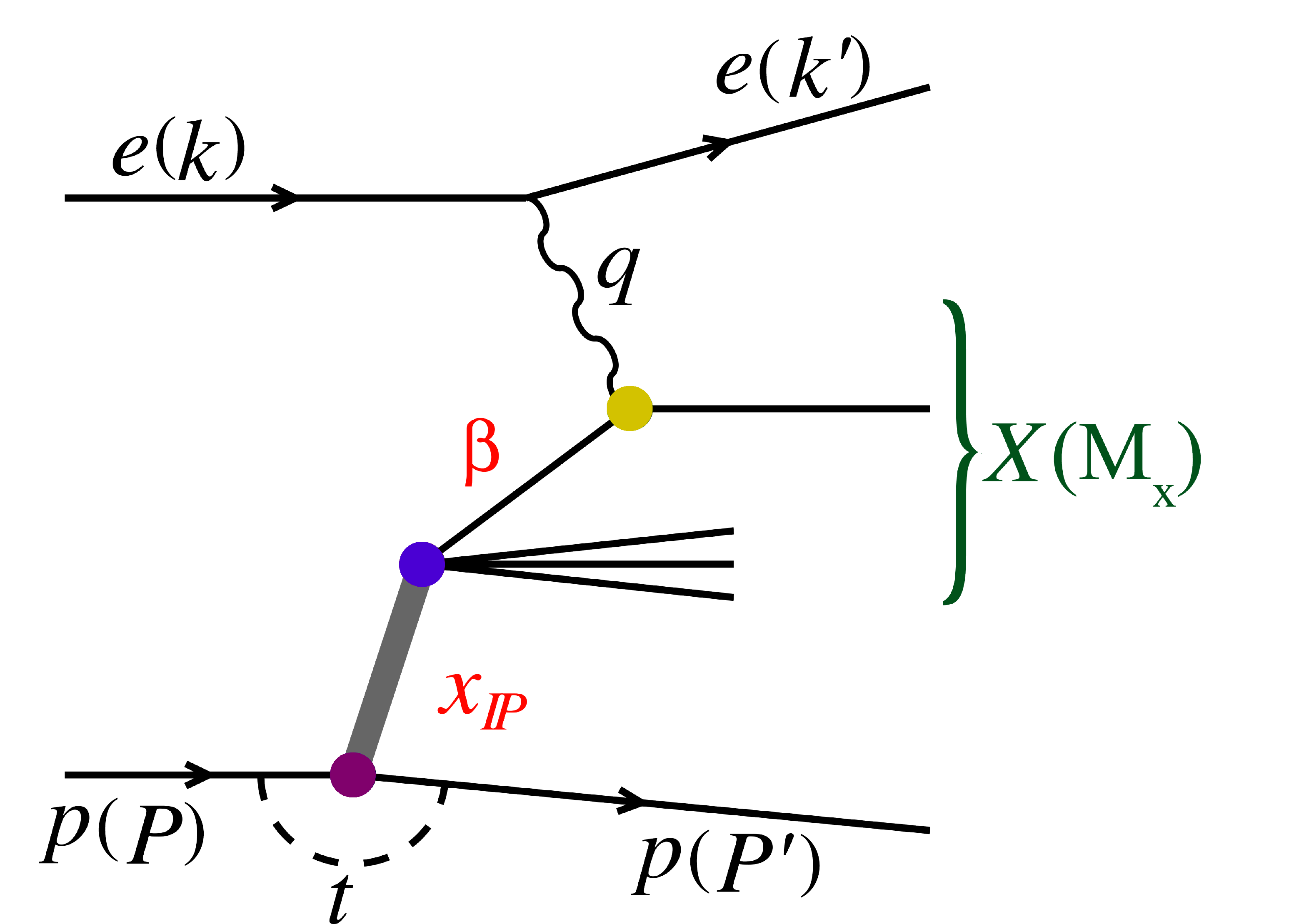}}  
\begin{center}
\caption{{\small A schematic parton model diagram of inclusive diffractive DIS $ep \rightarrow ep X$. Four-momenta are indicated in parentheses as well. The variable $\beta$ is the momentum fraction of the struck quark. } \label{fig:Feynman}}
\end{center}
\end{figure}

In this section, we review in details the theoretical formalism that describes diffractive DIS of charged leptons ($\ell^\pm$) of proton ($p$) at HERA.
As mentioned in the Introduction, the neutral current diffractive DIS process $ep \rightarrow ep X$, where $X$ represents the hadronic final state, is a DIS process in which a percent of the interacting proton remains intact. It has shown that, in $ep$ collisions at HERA, this hard diffraction contributes a fraction of order 8-10\% to the total DIS cross sections~\cite{Aktas:2006hy}. Such a process can be explained in terms of particle exchanges with the net vacuum quantum numbers, namely Pomeron and Reggeon, though there are different theoretical approaches to describe this process~\cite{Royon:2006by}. 
Fig.~\eqref{fig:Feynman} shows the schematic parton model diagram of inclusive diffractive DIS $e p \rightarrow e p X$. Four-momenta are indicated in parentheses as well. The variable $\beta$ is the momentum fraction of the struck quark.

Note that the hadronic final state $X$ separated from the intact proton by a rapidity gap (RG), so that the most momentum is carried by proton.
We start now by briefly reviewing the definition of the diffractive DIS reduced cross sections and of the associated kinematical variables which are relevant for the description of the diffractive lepton-proton ($\ell^\pm p$) scattering. As can be seen from Fig.~\eqref{fig:Feynman}, in addition to the common variables for describing DIS, which are the photon virtuality $Q^{2} = -q^{2} = -( k - k^{\prime})^{2}$, the longitudinal momentum fraction $x$ and the inelasticity $y$, one needs also some other essential variables for describing diffractive DIS. These new variables include: the mass of the diffractive final
state $M_{X}$ that is usually replaced by the Bjorken variable defined for the diffractive exchange $\beta$, the squared four-momentum transferred at the proton vertex $t$ is the square of the difference between the four-momentum of the outgoing and incoming protons, $P$ and $P^{\prime}$, respectively. The fraction of the momentum of proton carried by the diffractive exchange, which is denoted by $x_{\pom}$, is related to $x$ and $\beta$ variables through $x_{\pom} = x/\beta$.

The only available diffractive DIS datasets come from HERA $ep$ collider measurements that provided by the H1 and ZEUS Collaborations~\cite{Aktas:2006hy,Chekanov:2008fh,Aaron:2012hua,Aaron:2012ad,Chekanov:2009aa,Aktas:2006up,Aaron:2012zz,Aaron:2012md,Aktas:2007hn,Chekanov:2008cw, Chekanov:2005vv, Chekanov:2004hy}. These measurements are usually presented in terms of the so-called diffractive reduced cross sections $\sigma_r^{D(4)} {(\beta, Q^{2}; x_{\pom},t)}$  which depend on the diffractive structure functions, $F_2^{D(4)}$ and $F_L^{D(4)}$, according to the following equation,

\begin{align}
\label{eq:sigmar}  
\sigma_r^{D(4)}
{(\beta,
Q^{2};
x_{\pom}, t)}
=
&F_2^{D(4)}
{(\beta, Q^{2};
x_{\pom}, t)} 
\nonumber  \\
&-\dfrac
{y^{2}}
{1 + 
(1-y)^{2}}
F_L^{D(4)}
{(\beta,
Q^{2};
x_{\pom},
t)}\,.
\end{align}

In the experimental point of view, the longitudinal structure function, $F_L^{D(4)}$, is small enough and can be neglected in some specific limits where $y$ is not to close to unity. As can be easily checked, if one neglects $F_L^{D(4)}$, the diffractive reduced cross section $\sigma_{r}^{D(4)}$ will be approximately equal to $F_2^{D(4)}$. It is worth noting here that the diffractive DIS data are usually presented as $t$-integrated reduced cross section. For our analysis, we consider the diffractive reduced cross sections $\sigma_r^{D(4)}{(\beta, Q^{2}; x_{\pom}, t)}$ as presented in Eq.~\eqref{eq:sigmar} with the contribution of $F_L^{D(4)}$.

It is well established now, from the factorization theorem for diffractive DIS in perturbative QCD~\cite{Collins:1996fb,Collins:2001ga,Collins:1997sr}, that the diffractive structure functions can be written as a convolution of hard scattering coefficient functions $C_{2/L, i}$, with diffractive PDFs of flavour $i$; $f_{i}^{D}(z, Q^{2}; x_{\pom}, t)$,

\begin{align}
\label{eq:factorisation}
&F_{2/L}^
{D(4)}
(\beta,
Q^{2};
x_{\pom},
t) \nonumber\\
& \hspace{0.6cm} =
\sum_i
\int_{\beta}
^{1}
\frac{dz}{z} 
\, C_{2/L,
 i}
\Big(
\frac{\beta}{z}
\Big)\,
f_{i}^{D}
(z, 
Q^{2};
x_{\pom},
t)\,.
\end{align}

In above equation, the sum runs over all active partons including the quarks and gluons.
The coefficient functions, $C_{2/L,i}$, can be calculated perturbatively in QCD, whereas the diffractive parton densities, $f_{i}^{D}$, are nonperturbative quantities and should be determined through QCD global analyses of diffractive experimental datasets. 
It should be also noted that the diffractive PDFs, just like the usual parton densities in inclusive DIS, satisfy the well-know DGLAP evolution equations~\cite{Berera:1995fj,Martin:2006td,Kunszt:1996pj}. Moreover, one can use the same coefficient functions, $C_{2/L, i}$ as in inclusive DIS~\cite{Vermaseren:2005qc} for calculating diffractive structure functions of Eq.~\eqref{eq:factorisation}.

%
\subsection{ Quark and gluon diffractive PDFs }\label{sec:diffractive-PDFs}
%

According to the Regge factorization hypothesis, the diffractive PDFs can be separated into two terms. One term involving the $x_{\pom}$ and $t$ variables, represents the Pomeron and Reggeon fluxes, and the other term, including the leptonic variables $\beta$ and $Q^{2}$, describes the hard scattering of the photon with the partonic structure of the Pomeron and Reggeon. Hence, considering the Regge factorization scheme, the diffractive PDFs can be formulated as follows,

\begin{align}\label{eq:fD}
f_{i/p}
^D(\beta, 
Q^{2};
x_{\pom}, t)
= &f_{{\pom}/p}
(x_{\pom},
 t)
f_{i/{\pom}}
(\beta,
 Q^{2})
\nonumber\\
& + f_{{\reg}/p}
(x_{\pom},
 t)
f_{i/{\reg}}
(\beta,
Q^{2}) \,,
\end{align}
where the $f_{{\pom}/p}$ and $f_{{\reg}/p}$ are the Pomeron and Reggeon fluxes, respectively, and the $f_{i/{\pom}}$ and $f_{i/{\reg}}$ represent the parton densities of the Pomeron and Reggeon.
It should be note that one of the advantages of Regge factorization is the separation of the perturbative scale of the process, $Q^{2}$, from the $x_{\pom}$ behavior that is genuinely nonperturbative. However, the normalization of the Pomeron flux is ambiguous, since the Pomeron is not a particle. In that way, the separation of the flux from the Pomeron density will be quite arbitrary.

As mentioned before, the dependence of the quark and gluon density functions of the Pomeron, $f_{i/{\pom}} (\beta, Q^{2})$, on the $Q^{2}$ scale can be obtained by the standard DGLAP evolution equations, provided their dependence on $\beta$ is determined at an initial scale $Q_{0}^{2}$. 
Since the HERA diffractive DIS datasets could only constrain the sum of diffractive PDFs, 
and on the other hand, the available data are not sufficient enough to constrain all shape parameters of the separate flavors, diffractive PDFs are usually parameterized as simple functional forms at the initial scale in terms of quark $zf_{q}(z, Q_{0}^{2}) $ and gluon $zf_{g}(z, Q_{0}^{2})$ distributions. The quark and antiquark distributions are assumed to be equal, $f_{u} = f_{d} = f_{s} = f_{\bar{u}} = f_{\bar{d}} = f_{\bar{s}}$.

It should be also noted that, $z$ is the longitudinal momentum fraction of the struck parton
with respect to the diffractive exchange that differs to $\beta$ when the higher-order processes are also included. In the present work, the Pomeron partonic densities are parameterized at the initial scale $Q_{0}^{2} = 1.8$ $\gev^{2}$ as follows,

\begin{align}
&&\label{eq:densityquark}
zf_q (z,
Q_{0}^{2})
=\alpha_{q} \,
z^{\beta_{q}}
(1-z)^{\gamma_{q}}
(1 + \eta_{q}
\sqrt{z}),   \\
&& \label{eq:densitygloun}
zf_g(z, Q_{0}^{2}) =
\alpha_{g}\, z^{\beta_{g}}
(1-z)^{\gamma_{g}}
(1+\eta_{g}\sqrt{z}).
\end{align}

One should notice here that an extra factor $\exp[-0.001/(1-z)]$ is simply multiplied to the above parameterizations, in order to ensure that they go to zero for $z \to 1$.
Considering the above parameterizations, the parameters $\gamma_{q}$, $\gamma_{g}$, $\eta_q$ and $\eta_g$ have the freedom in our analysis to extract from the QCD fit, so that can get negative or positive values.
Such parameterizations form have been used in several analyses~\cite{Aktas:2006hy,Chekanov:2009aa,Goharipour:2018yov}, and its validity has experimentally been tested by HERA $ep$ experiments. For the case of Reggeon partonic densities in Eq.~\eqref{eq:fD}, we use the parameterizations forms of the NLO GRV group which have been obtained from a QCD analysis of the pion structure functions data, see Ref.~\cite{Gluck:1991ng} for details.

The dependence of the diffractive PDFs $f_{i}^{D}(\beta, Q^{2}; x_{\pom}, t)$ introduced in Eq.~\eqref{eq:fD} to $x_{\pom}$ is given by the flux factors of the Pomeron and Reggeon.
In the present study, we use the same functional form as in Refs.~\cite{Aktas:2006hy,Chekanov:2009aa,Goharipour:2018yov},  

\begin{align}
\label{eq:flux}
f_{\pom, \reg}
(x_{\pom}, t)
= A_{\pom, \reg}
\,\frac{e^{B_{\pom,
\reg} \, t}}
{x_{\pom}^{2
\alpha_{\pom, \reg}
(t)-1}}\,,
\end{align}
where $\alpha_{\pom, \reg}(t)$ is considered to be a linear function in terms of $t$, $\alpha_{\pom, \reg}(t) = \alpha_{\pom,\reg}(0)+{\alpha}_{\pom, \reg}^{\prime} t$. Hence, the Reggeon normalization factor, $A_{\reg}$, and also the Pomeron and Reggeon intercepts, $\alpha_{\pom}(0)$ and $\alpha_{\reg}(0)$, are free parameters in our analysis and should be extracted from QCD fit to diffractive DIS datasets. Note that, according to Eq.~\eqref{eq:fD} for diffractive PDFs, the value of Pomeron normalization parameter, $A_{\pom}$, is absorbed in $\alpha_{q}$ and $\alpha_{g}$ parameters. The other parameters involving in Eq.~\eqref{eq:flux} are considered to be constant with the values given in Ref.~\cite{Goharipour:2018yov}.

%
\subsection{ Heavy quark contributions and numerical calculations }\label{sec:Heavy-quark}
%

It is worth noting that the evolution of diffractive PDFs is performed using the publicly available {\tt APFEL} packag~\cite{Bertone:2013vaa}. Note also that, like for the case of {\tt ZEUS-2010-DPDFs}~\cite{Chekanov:2009aa} and {\tt GKG18}~\cite{Goharipour:2018yov} analyses, the heavy quark contributions to the structure functions are considered using GM-VFNS~\cite{Harland-Lang:2014zoa,Thorne:1997ga,Thorne:2006qt}.
Since this analysis is based on the {\tt APFEL} packag~\cite{Bertone:2013vaa}, we specifically use the {\tt FONLL-B} GM-VFNS scheme~\cite{Forte:2010ta} for our NLO QCD fits which implemented in this package. 
Our GM-VFNS has a maximum of $N_f = 5$ active quarks, and for the case of heavy quark masses, we fix the charm and bottom quark masses at $m_{c}=1.40$ GeV  and $m_{b}=4.75$ GeV, respectively. As a last point, it should be mentioned that the value of strong coupling constant at the $Z$-boson scale is considered to be $\alpha_{s}(M_{Z}) = 0.1176$ consistent with the PDG average~\cite{Tanabashi:2018oca} and with very recent high precision determinations~\cite{Bruno:2017gxd,Verbytskyi:2019zhh,Ball:2018iqk,Zafeiropoulos:2019flq}.
In the present study, by performing a QCD analysis of the diffractive DIS data from HERA measurements in the presence of heay quark contributions, we obtain diffractive PDFs and their uncertainties at the NLO accuracy in QCD. To this aim, we use the CERN program library {\tt MINUIT}~\cite{James:1975dr} for performing fit procedure and determining the unknown parameters.

%
\subsection{ Higher-twist effects }\label{sec:Higher-twist}
%

As we discussed earlier, in this article, we plan to show that the diffractive DIS data at low values of $\gev^2$ could provide the first evidence for the HT effects in diffractive DIS in the perturbative domain, and hence, could open a possibility for further theoretical and experimental investigations in such high energy processes. In this section, our aim is to introduce the nonperturbative HT corrections to diffractive structure functions in the diffractive DIS experiment at HERA. 
In a wide kinematic region in terms of $x$ and $Q^{2}$, one can describe the structure functions of the DIS using leading-twist (LT) corrections in QCD. However, for small values of $Q^{2}$ and large values of $x$, the structure functions in Eq.~\eqref{eq:sigmar} should be corrected for some nonperturbative corrections. In general, in such region, two types of corrections should be considered which are the target mass corrections (TMCs) and the higher-twist (HT) effects. In the case of TMCs, structure functions are corrected by modifying some quantities and adding new sentences, which can be found in Ref.~\cite{Schienbein:2007gr} for full study in this area.
For HT case, it is customary to correct the leading-twist structure function by adding a sentence that is inversely related to $Q^{2}$. More precisely, the higher-twist effects parameterize in the form of a phenomenological unknown function, and then they obtain the values of the unknown parameters using the fitting to the experimental data~\cite{Accardi:2016qay}.

The phenomenological form for the HT effects in the corrected structure function is considered as follows
	
\begin{align}\label{equ:equ7}
F_{2} (x, Q^{2}) = 
F_{2}^{LT}
(x,Q^{2})
\left(1
+\frac{C_{HT}(x)}
{Q^{2}} \right) \,,
\end{align}
where $F_{2}^{LT}$ represents the leading-twist structure function, and the higher-twist coefficient function is parameterized as follows

\begin{align}\label{equ:8}
C_{HT}(x) = 
h_{0} \, x^{h_{1}}
(1 + h_{2} x) \,. 
\end{align}
The $h_{0}$ parameter represents the higher-twist correction general scale, while $h_{1}$ controls the increase of the $C(x)$ coefficient at large $x$, and the parameter $h_{2}$ allows the probability of a higher-twist at small values of momentum fraction $x$. In Eq.~\eqref{equ:8}, $h_i \, \, \{i=0,1,2\}$ should be determined along with the fit parameters and then keep fixed. 

To conclude this section, we should mentioned here that we follow a standard method to consider the higher twist effects in our QCD analysis of fully inclusive DIS structure functions. A a result of this study, we will show that such corrections are sizable at large values of the Bjorken variable $x$ and small region of $Q^{2}$. Detailed studies presented in Refs.~\cite{GolecBiernat:2007kv,GolecBiernat:2009zza,GolecBiernat:2001mm} indicate that, in addition to this standard higher twist contribution, one could also consider the twist-$4$ contribution which dominates in the region of large $\beta$. This contribution comes from the diffractive production of the $q \bar q$ pair from the longitudinally polarized virtual photons ($L q \bar q$) in diffractive DIS. Formally, this contribution suppress by a power of $1/Q^{2}$ with respect to the leading twist-$2$ contribution and is particularly important for the longitudinal diffractive structure function $F_L^D$. The results presented in Ref.~\cite{GolecBiernat:2007kv} revealed that the diffractive gluon distribution obtained through a QCD fit with twist-$4$ has a stronger peaked near large $\beta$. The longitudinal structure functions for the large value of $\beta \sim 1$ is also dominated by the twist-$4$ contribution.
Finally, we should note here that the scope of the present study is limited to the standard higher twist effects. The kinematical cuts that we applied on the datasets in our analysis are discussed in Sec.~\ref{sec:Kinematical-cuts}. For the $\beta$, we apply $\beta \leq 0.80$ over all datasets used in this study and hence such correction can be strongly reduced and safely ignored, thus they are not considered.

%
\section{ Difrractive DIS datasets }\label{sec:Difrractive-DIS-data-sets}
%

This section includes a comprehensive introduction to the diffractive DIS experimental datasets that we are used in our diffractive PDFs analysis, discussing also the kinematical cuts and the treatment of experimental uncertainties.

%
\subsection{ Experimental data }\label{sec:Experimental-data}
%

\begin{figure}[t!]
\vspace{0.20cm}
\resizebox{0.55\textwidth}{!}{\includegraphics{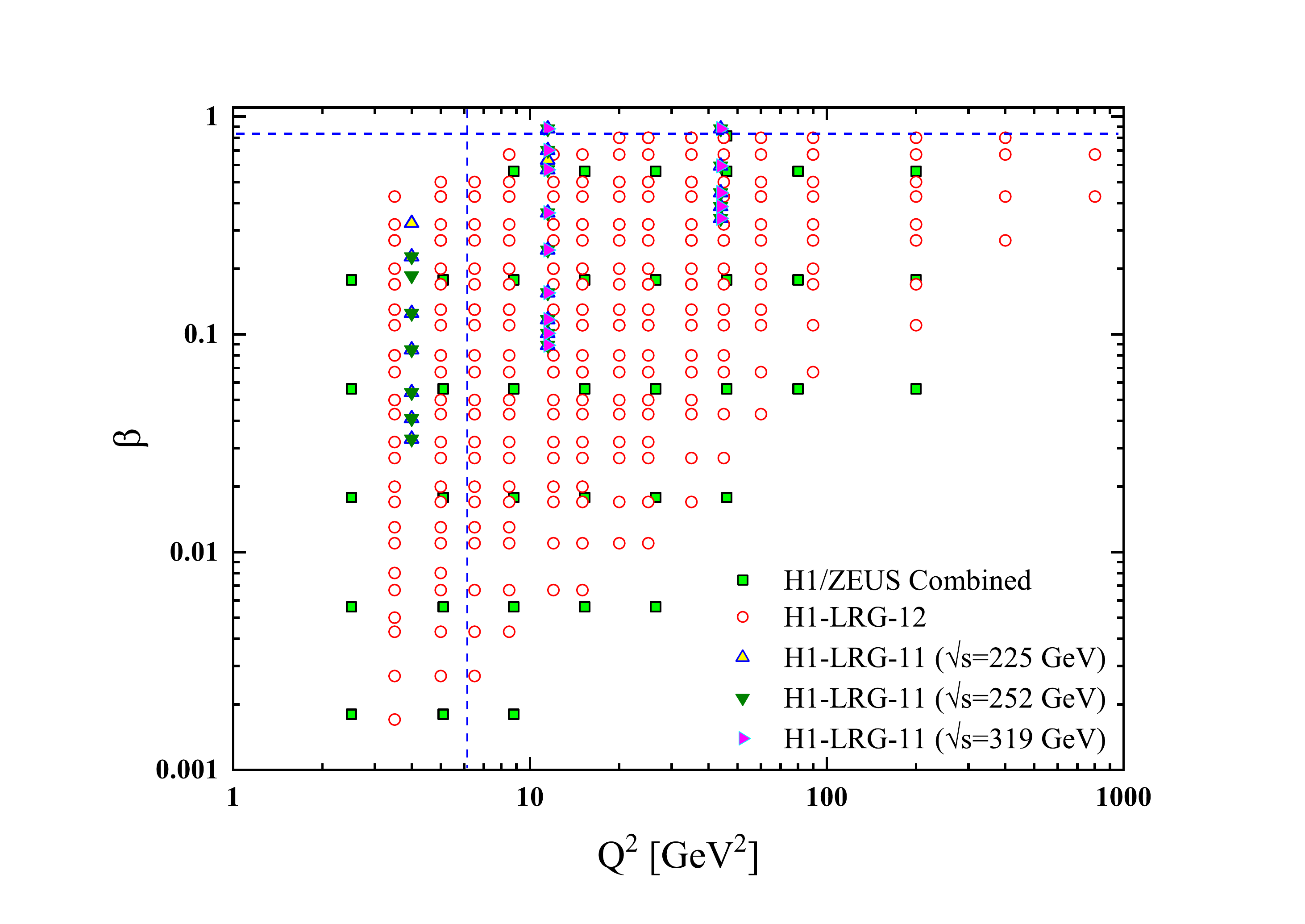}}  
\begin{center}
\caption{{\small Different experiments of diffractive DIS datasets in the $\beta$ and $Q^2$ plane. The dashed lines represent the kinematic cuts applied on $Q^2$ and $\beta$ in this analysis. The data points lying outside these lines shown in the figure are only excluded in the present QCD fits. } \label{fig:DataxQ}}
\end{center}
\end{figure}

In this analysis, we present all available inclusive diffractive DIS measurements on lepton-proton ($\ell p$) scattering at HERA.
In Fig.~\eqref{fig:DataxQ}, we show the kinematical coverage in the $\beta$ and $\gev^2$ plane for the different experiments of diffractive DIS $ep$ scattering datasets used in this analysis. The dashed line represents the kinematic cuts applied on $\gev^2$ and $\beta$. The data points lying outside these lines shown in the figure are only excluded in the present QCD fits.

For our analysis, in particular, we use LRG data from the H1 Collaboration~\cite{Aaron:2012zz,Aaron:2012ad} as well as the most recent data from H1/ZEUS combined inclusive diffractive cross section measurements~\cite{Aaron:2012hua}.

\begin{table*}[htb]
	\caption{\small List of all diffractive DIS data points used in our global analysis. For each
		dataset we have provided the corresponding publication reference, the kinematical coverage of $\beta$, $x_{\pom}$, and $Q^{2}$. We have also displayed the number of data points before the baseline kinematical cuts. The details of kinematic cuts imposed on these data have been explained in details in the text. } \label{tab:DDISdata}
	\begin{tabular}{l  c  c  c  c  c  c c }
		Experiment & Observable & [$\beta^{\rm{min}}, \beta^{\rm{max}}$] & [$x_{\pom}^{\rm{min}}, x_{\pom}^{\rm{max}}$]  & $\gev^{2}\,[{\text{GeV}}^2]$  & \# of points & Reference
		\tabularnewline
		\hline\hline
		H1-LRG-11 $\sqrt{s} = 225$ GeV & $\sigma_r^{D(3)}$ & [$0.033$--$0.88$]    & [$5\times 10^{-4}$ -- $3\times10^{-3}$] & 4--44 & \textbf{22}  & \cite{Aaron:2012zz}  \\
		H1-LRG-11 $\sqrt{s} = 252$ GeV & $\sigma_r^{D(3)}$ & [$0.033$--$0.88$]    & [$5\times 10^{-4}$ -- $3\times10^{-3}$] & 4--44 & \textbf{21}  &  \cite{Aaron:2012zz} \\
		H1-LRG-11 $\sqrt{s} = 319$ GeV & $\sigma_r^{D(3)}$ & [$0.089$--$0.88$]    & [$5\times 10^{-4}$ -- $3\times10^{-3}$] & 11.5--44 & \textbf{14} &  \cite{Aaron:2012zz}   \\	
		H1-LRG-12 & $\sigma_r^{D(3)}$ & [$0.0017$--$0.80$]   & [$3\times10^{-4}$ -- $3\times10^{-2}$] & 3.5--1600 & \textbf{277}   &  \cite{Aaron:2012ad}  \\	
		H1/ZEUS combined & $\sigma_r^{D(3)}$  &   [$0.0018$--$0.816$]   & [$3\times10^{-4}$ -- $9\times10^{-2}$] & 2.5--200 & \textbf{192}  & \cite{Aaron:2012hua}   \\			
		\hline \hline
		\multicolumn{1}{c}{\textbf{Total data}} ~~&~~ &~~ &~~& ~~&~~\textbf{526}  \\  \hline
	\end{tabular}
\end{table*}

In the following, we discuss each of these measurements in more details.
In addition to the Fig.~\ref{fig:DataxQ}, a full list of all datasets has been presented in Table~\ref{tab:DDISdata} as well. Note that for each dataset listed in this table, we have provided the related published references, the kinematic intervals of $\beta$, $x_{\pom}$ and $ Q^2 $ variables, and also the number of data points. As one can see, the included datasets cover a wide range of $\beta$ and $Q^{2}$, though they belong to small values of $x_{\pom}$. It should be also noted that, one needs to impose some kinematic cuts in order to avoid nonperturbative effects and also possible problems with the chosen theoretical framework. In this way, the total number of data points that are finally included in the analysis will be decreased after imposing related cuts. We discuss this issue in details at the beginning of the next section, where we investigate the appropriate value of the minimum cut on $\gev^{2}$ to consider the HT effects.

We have used three different datasets in this analysis.
The latest H1/ZEUS combined dataset for the reduced diffractive cross sections, $\sigma^{D(3)}_{r}(ep \rightarrow {epX})$~\cite{Aaron:2012hua} has been used. This calculations used samples of diffractive DIS $ep$ scattering data at a center-of-mass energy of $\sqrt{s} = 318$  {GeV} at HERA collider. This precise measurement covers range of photon virtualities $2.5 \gev^{2}$ to $200 \gev^{2}$ and $0.0018 \leq \beta \leq 0.816$. It should be noted that the most recent analyses from {\tt GKG18-DPDF}~\cite{Goharipour:2018yov}, {\tt HK19-DPDF}~\cite{Khanpour:2019pzq} and the analysis in Ref.~\cite{Ceccopieri:2016rga}, used the H1/ZEUS combined datasets in their analyses, however, those studies make no attempt to consider the beneficial effect arising from the inclusion of HT terms. 

Another dataset is the measurement of inclusive diffractive DIS from H1-LRG-11, which is derived from the H1 detector in 2006 and 2007. These data correspond to three different center-of-mass energies namely $\sqrt{s}=225$, 252  and 319 GeV~\cite{Aaron:2012zz}. In these measurements, the reduced cross sections have been calculated in the range of photon virtualities $4 \gev^{2} \leq Q^{2} \leq 44 \gev^{2}$ for the center of mass $\sqrt{s}=225, 252 \gev$, and $11.5 \gev^{2} \leq Q^{2} \leq 44 \gev^{2}$ for the center-of-mass of $\sqrt{s}=319 \gev$. The masses of hadronic final state are in the range of $1.25 \leq M_{X} \leq 10.84$  and the proton vertex is considered to be $\vert{t}\vert < 1 \gev^{2}$. The diffractive variables are taken in the range of $5 \times 10^{-4} \leq x_{\pom} \leq 3 \times 10^{-3}$, $0.033 \leq \beta \leq 0.88$  and $0.089 \leq \beta \leq 0.88$ for the center-of-mass energies of $\sqrt{s}=225$, 252 and =319 $\gev$, respectively. 
 
Finally, we use in this QCD fit the H1-LRG-12 data~\cite{Aaron:2012ad}, where the diffractive scattering $e p  \rightarrow {e X Y}$ are corrected to the region $M_{Y} < 1.6 \gev$, and momentum transfer of $\vert{t} \vert < 1 \gev^{2}$. They span the wide $x_{\pom}$ range  $0.0003 \leq x_{\pom} \leq 0.03$ and covers the ranges of $1.8 \times 10^{-3} \leq \beta \leq 0.88$ in $\beta$ and  $3.5 \leq Q^{2} \leq 1600 \gev^{2}$ in photon virtuality and $1.11 \leq M_{X} \leq 48.99$.

%
\subsection{ Kinematical cuts }\label{sec:Kinematical-cuts}
%

In this section, we briefly review the kinematical cuts applied on the diffractive DIS datasets analyzed in this study. 
In Fig.~\ref{fig:DataxQ}, we shown the kinematical coverage in the $(\beta; Q^2)$ plane of the diffractive DIS data included in our QCD fits.
However to minimize the contamination from low-scale non-perturbative corrections such as the target mass corrections and higher-twist effects, one needs to impose some certain kinematical cuts on the $Q^{2}$ and the invariant final state mass $M_X$. Applying these sort of kinematical cuts could deserve a number of precise investigation. The approach used in this investigation is similar to that used by other analysis in literature (see for example Refs.~\cite{Goharipour:2018yov,Aktas:2006hy,Chekanov:2009aa,Khanpour:2019pzq} for clear review). 

Like for the case of {\tt H1-2006}~\cite{Aktas:2006hy}, {\tt ZEUS-2010}~\cite{Chekanov:2009aa}, {\tt GKG18}~\cite{Goharipour:2018yov} and {\tt HK19-DPDF}~\cite{Khanpour:2019pzq} QCD fits, we apply a cut on $M_X$, $\beta$ and $Q^{2}$. To determine our diffractive PDFs from the QCD fit, we apply $\beta \leq 0.80$ over all datasets used in this study. The data point with $M_X < 2 \, {\text{GeV}}$ are excluded from the fit.

\begin{figure}[htb]
\vspace{0.20cm}
\resizebox{0.550\textwidth}{!}{\includegraphics{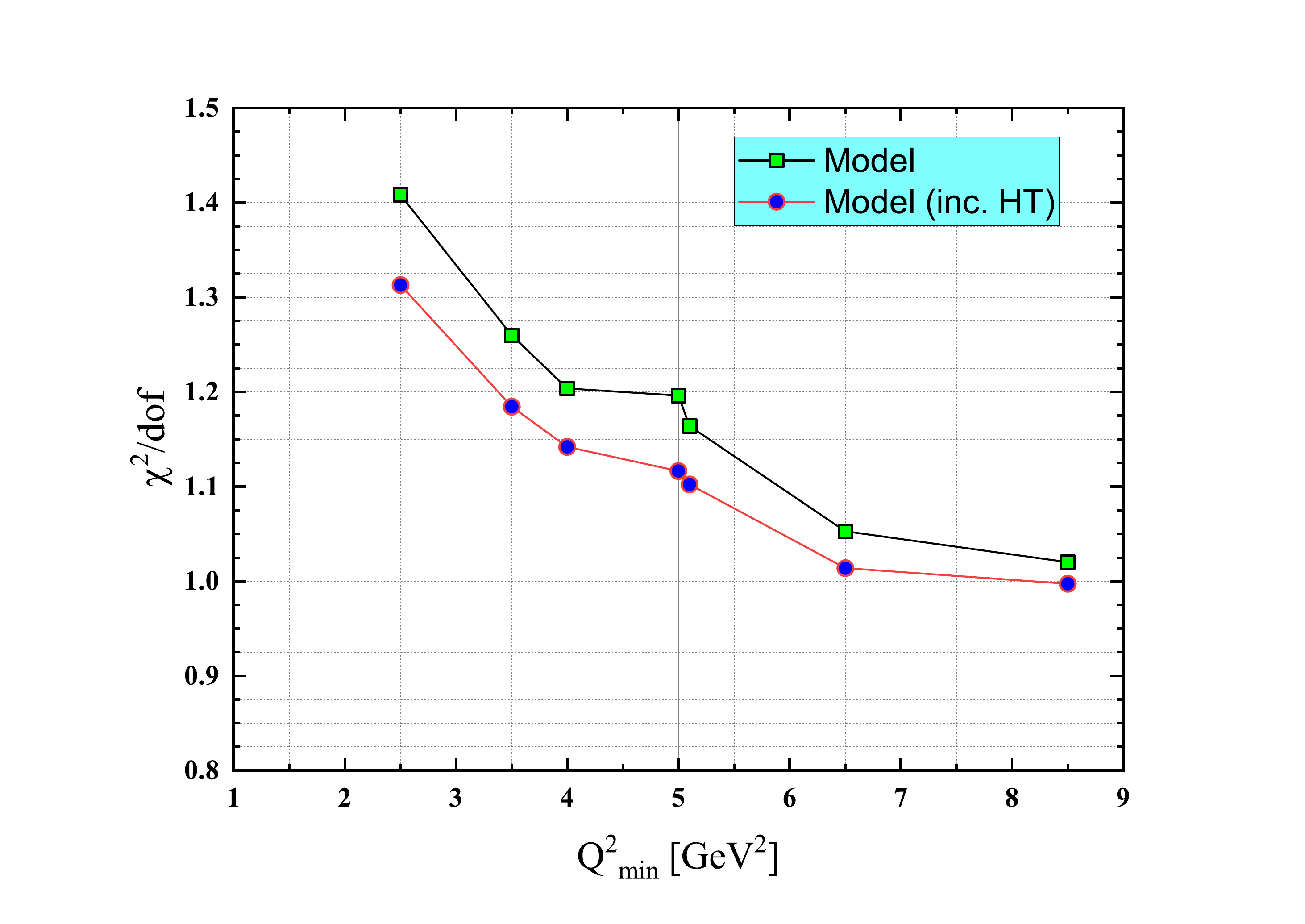}} 
\begin{center}
\caption{{\small  The value of the total $\chi^2/{\rm dof}$ vs the minimum $\gev^{2}$ of data, Q$^2_{\rm min}$, for all datasets entering in this study which represents our specific choice of the kinematical cuts on the Q$^2_{\rm min}$. } \label{fig:X22}}
\end{center}
\end{figure}

However the kinematical cut on the $Q^{2}$ needs some more discussions. To this end, the dependence of the $\chi^2$ on the $Q^{2}$ cut applied to the analyzed datasets is investigated in our analysis, and the results are shown in Fig.~\ref{fig:X22}. This figure shows the value of $ \chi^2 $ divided to number of degrees of freedom, $\chi^2/{\rm dof}$, vs the minimum $\gev^{2}$ of data included, $Q^2_{\rm min}$, for both of our QCD fits. Fig.~\ref{fig:X22} interestingly shows that the $\chi^2/{\rm dof}$ can be improved by the inclusions of the HT effects. In addition, as one can conclude from this figure, by increasing $Q^{2}$ no improvement can be seen for $\chi^2/{\rm dof}$. As shown in this plot, a significant decrease in $\chi^2/{\rm dof}$ can be seen at $Q^2_{\rm min} \leq 6.5 \, \gev^{2}$, so that the value of $\chi^2/{\rm dof}$ be around the unity. Hence, we prefer to consider $Q^2_{\rm min}$ = 6.5  $\gev^{2}$ as a best cut on $Q^{2}$. It is worth mentioning here that, in comparison to all other analyses in literature which consider 8.5 $\gev^{2}$ in their QCD fits, our assumption enables us to use much more data points in our analysis. In fact the total number of data points included in the analysis after imposing kinematical cuts is 499 which shows that, according to Table~\ref{tab:DDISdata}, only 27 data points are excluded.

\begin{figure*}[htb]
	\vspace{0.20cm}
	\resizebox{0.48\textwidth}{!}{\includegraphics{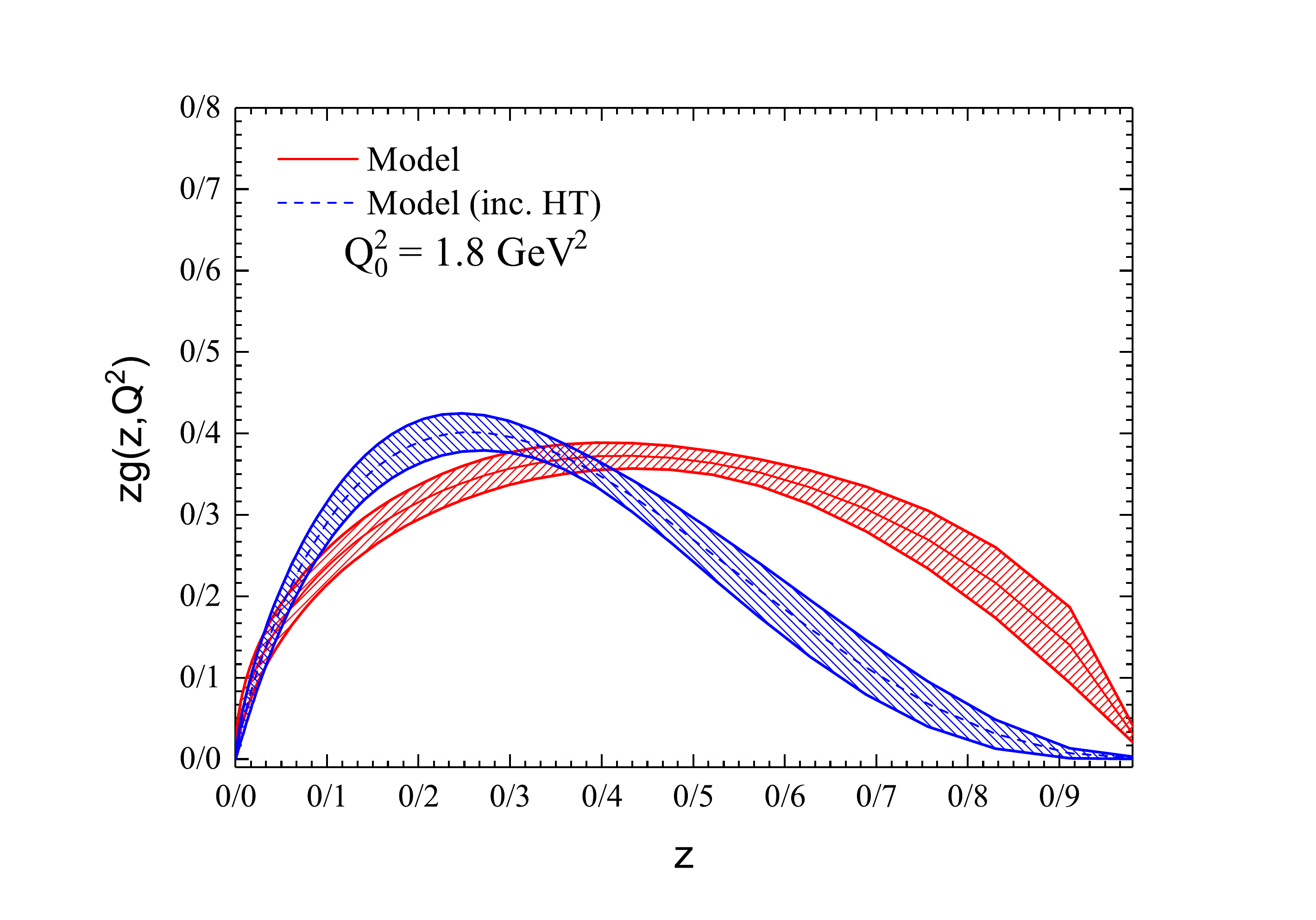}}  
	\resizebox{0.48\textwidth}{!}{\includegraphics{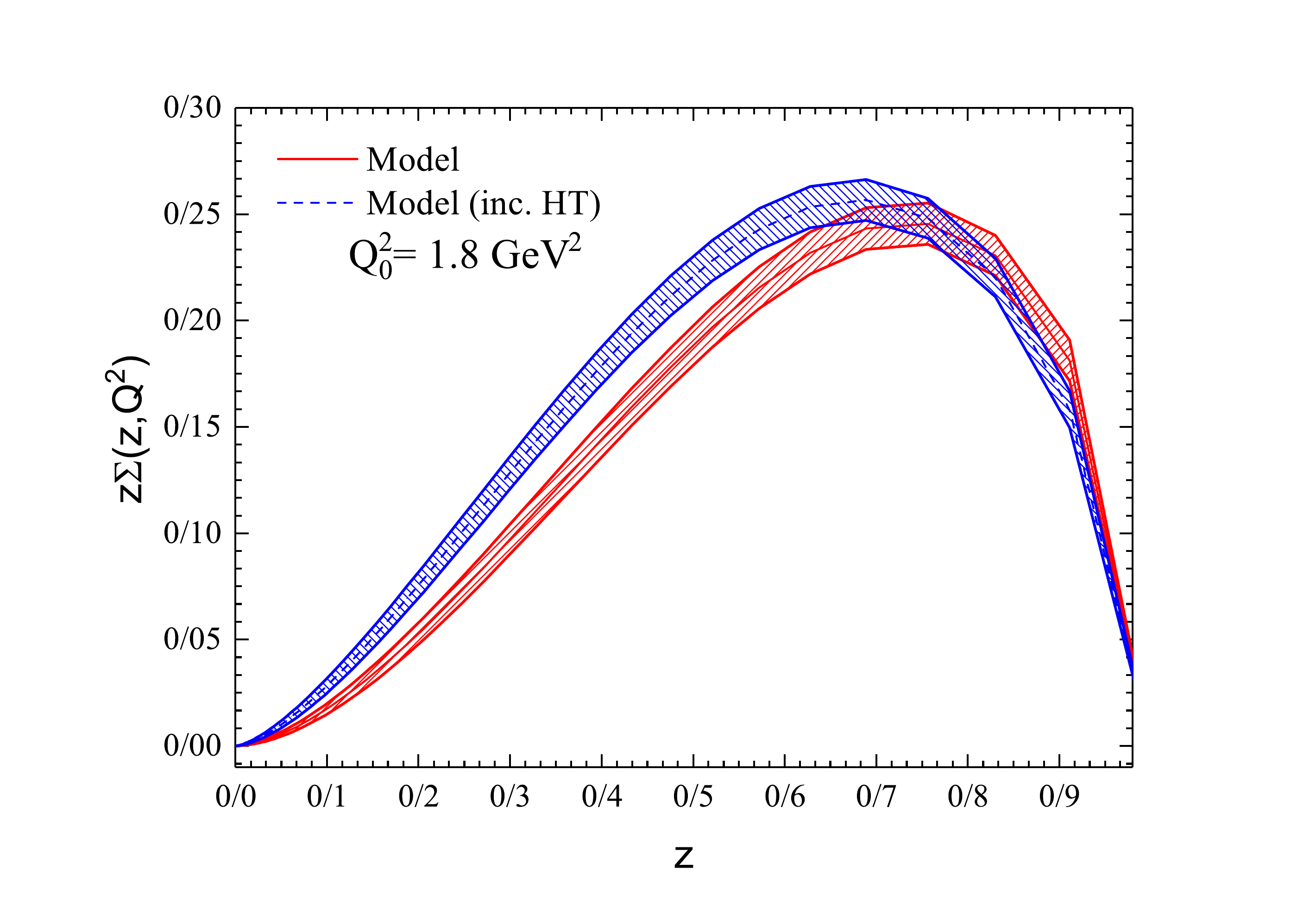}}
	\begin{center}
		\caption{{\small The diffractive PDFs as a function of momentum fraction $z$ obtained at the input scale of $Q_0^2$ = 1.8 GeV$^2$ for  both of our analyses with and without the HT corrections. The error bands represent the uncertainty estimation coming from the experimental errors. } \label{fig:DPDF-Q0}}
	\end{center}
\end{figure*}

\begin{figure*}[htb]
	\vspace{0.20cm}
	\resizebox{0.48\textwidth}{!}{\includegraphics{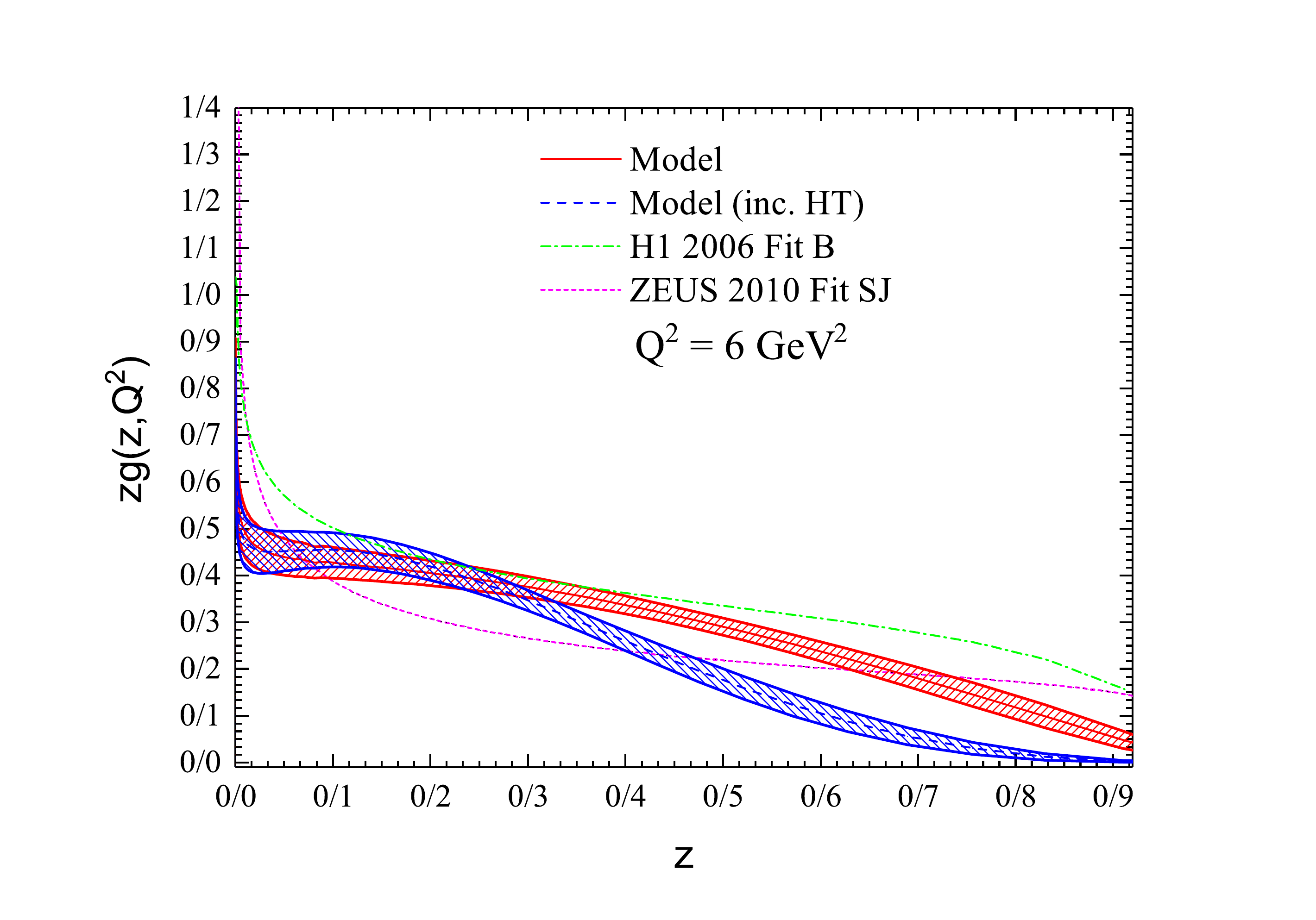}} 
	\resizebox{0.48\textwidth}{!}{\includegraphics{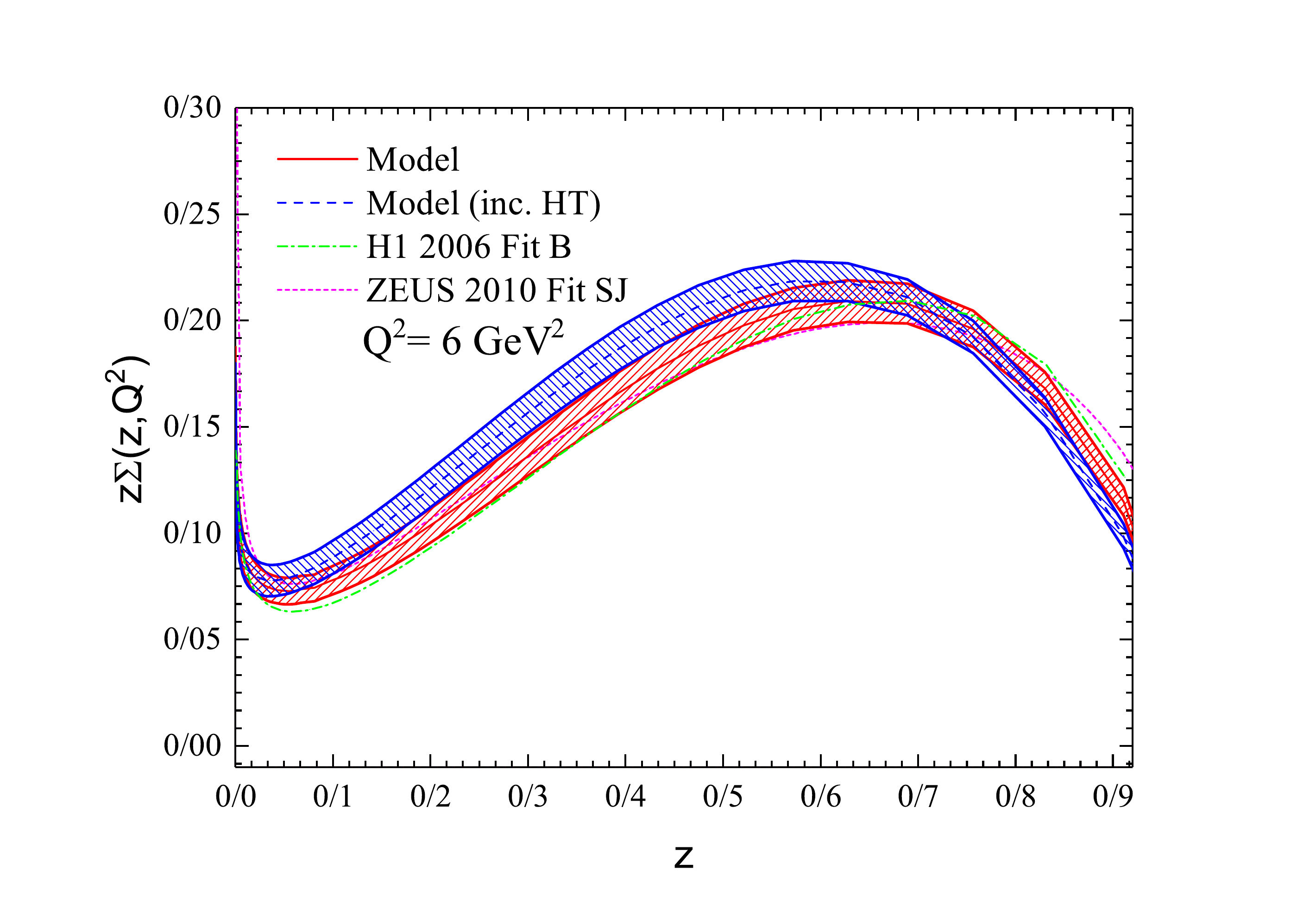}}
	\resizebox{0.48\textwidth}{!}{\includegraphics{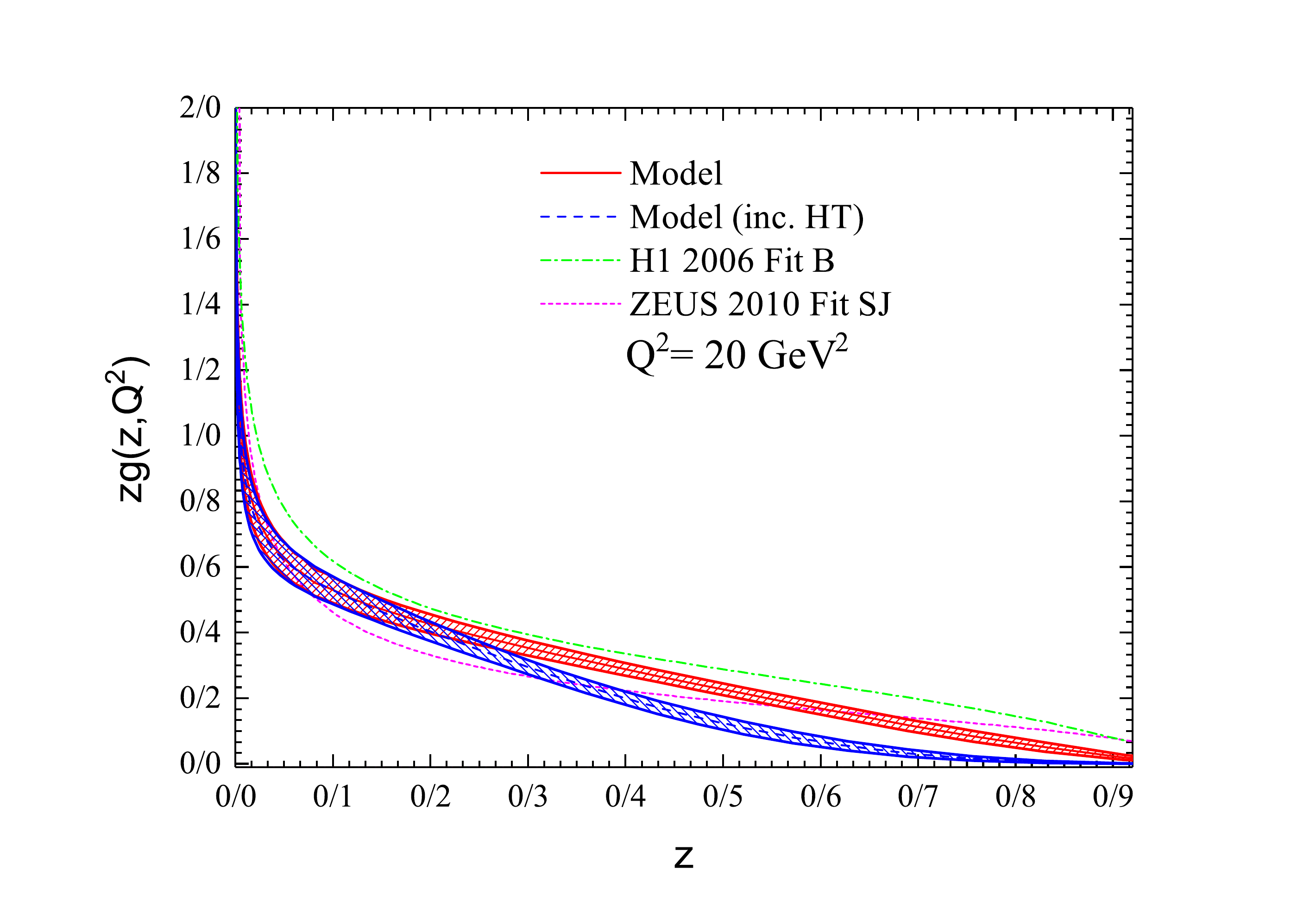}}  
	\resizebox{0.48\textwidth}{!}{\includegraphics{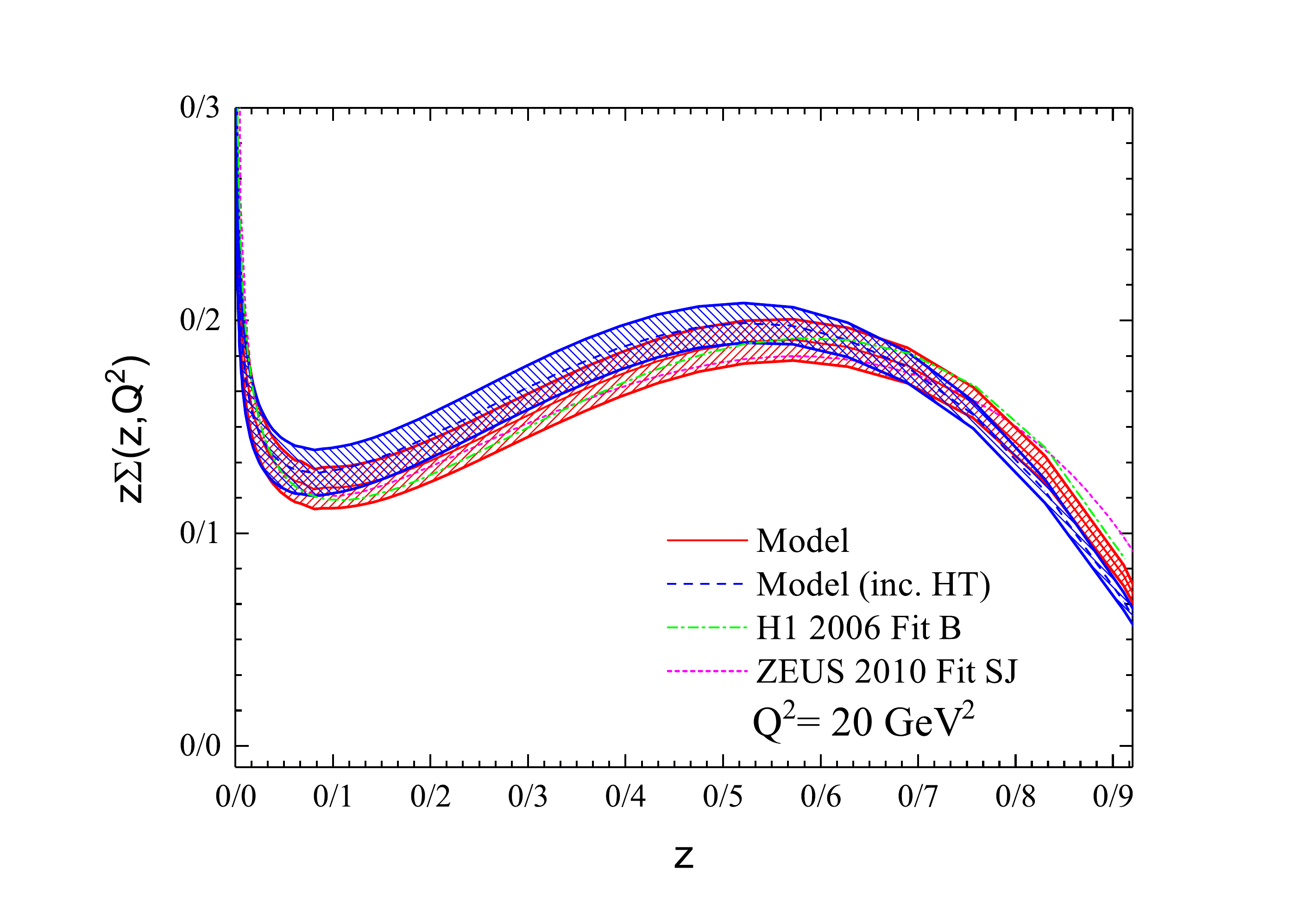}}
	\resizebox{0.48\textwidth}{!}{\includegraphics{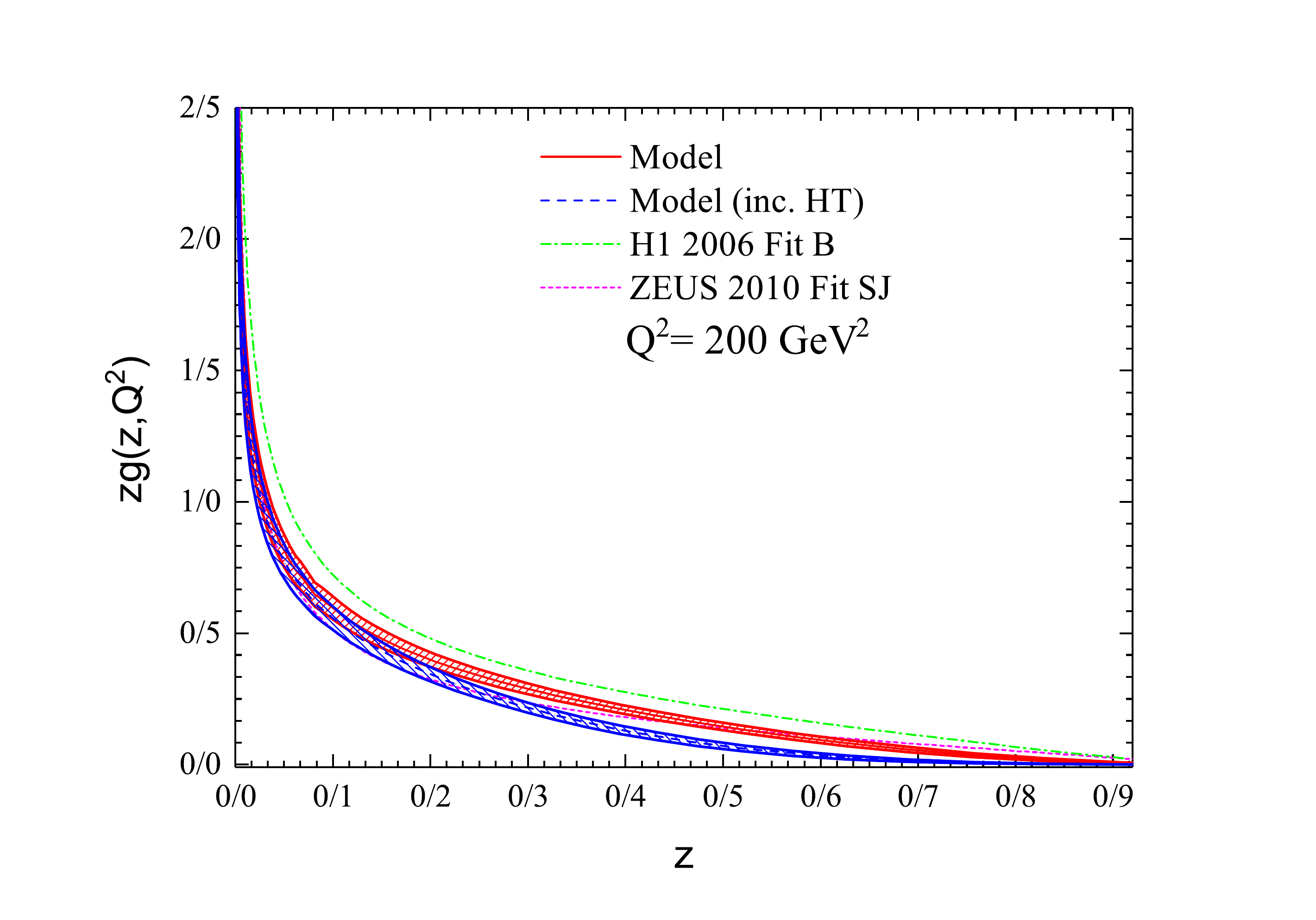}}  
	\resizebox{0.48\textwidth}{!}{\includegraphics{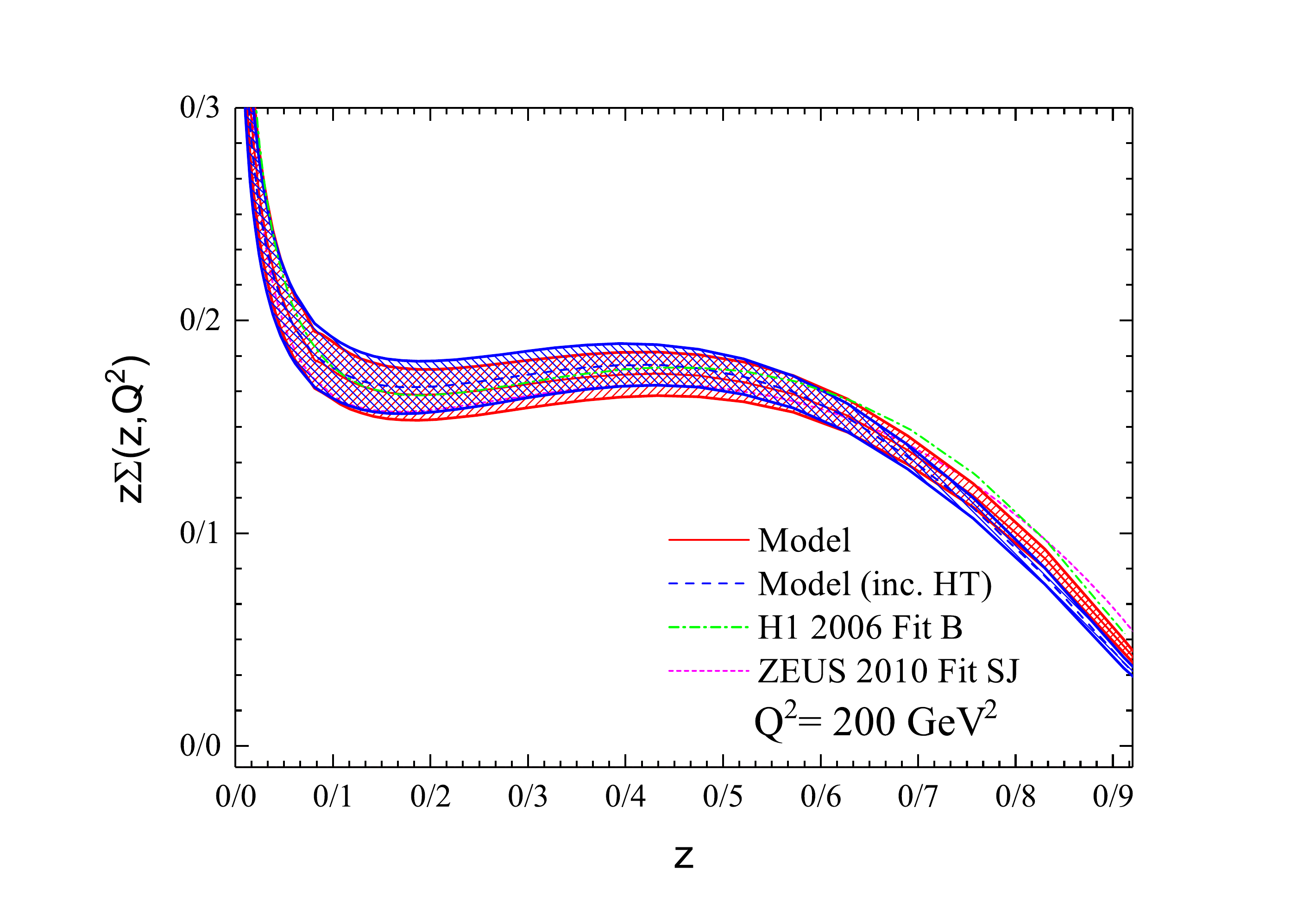}}
	\begin{center}
		\caption{{\small  The extracted diffractive PDFs for gluon and quark densities in three photon virtuality of $Q^2 = 6\,, 20$ and $200$ GeV$^2$ compared to the results of {\tt H1-2006 Fit B}~\cite{Aktas:2006hy} and {\tt ZEUS-2010 Fit SJ}~\cite{Chekanov:2009aa}. } \label{fig:DPDF-Q6-20-200}}
	\end{center}
\end{figure*}

\begin{figure*}[htb]
	\vspace{0.20cm}
	\resizebox{0.48\textwidth}{!}{\includegraphics{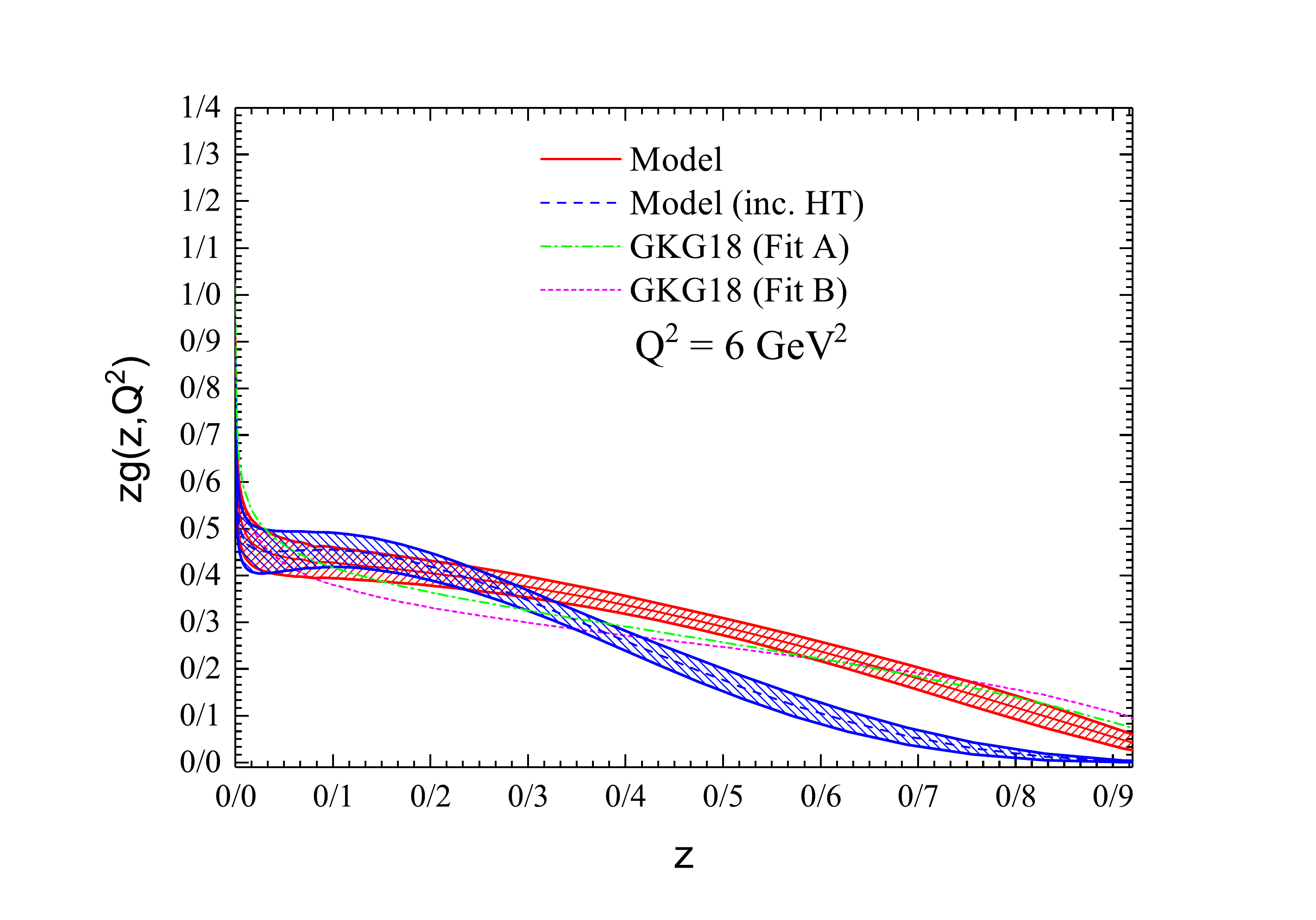}}  
	\resizebox{0.48\textwidth}{!}{\includegraphics{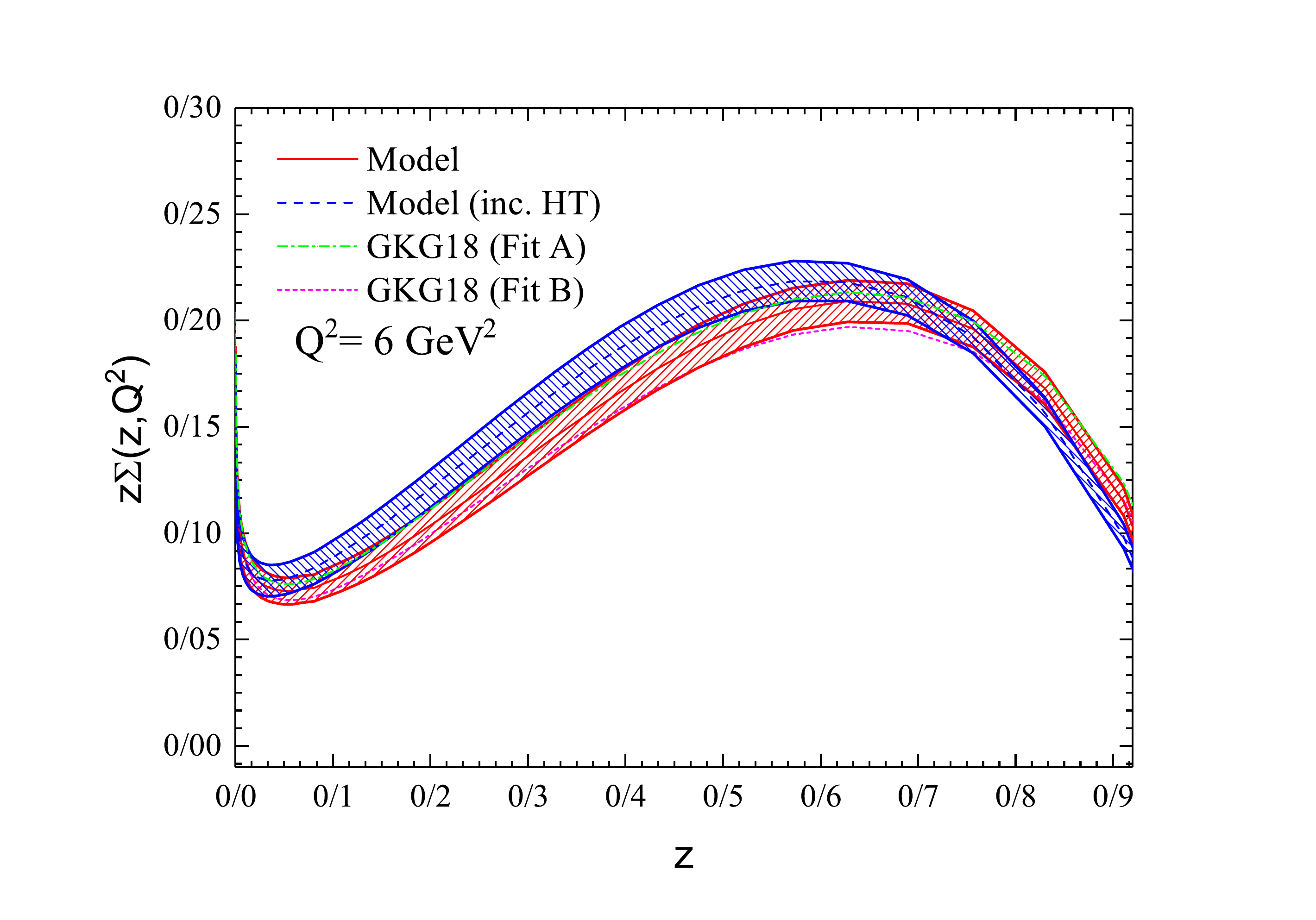}}
	\begin{center}
		\caption{{\small The extracted diffractive PDFs for gluon and quark densities at $Q^2 = 6$ $\gev^{2}$ compared to the most recent analysis by {\tt GKG18}~\cite{Goharipour:2018yov}. } \label{fig:GKG}}
	\end{center}
\end{figure*}

\begin{figure*}[htb]
	\vspace{0.50cm}
	\resizebox{0.950\textwidth}{!}{\includegraphics{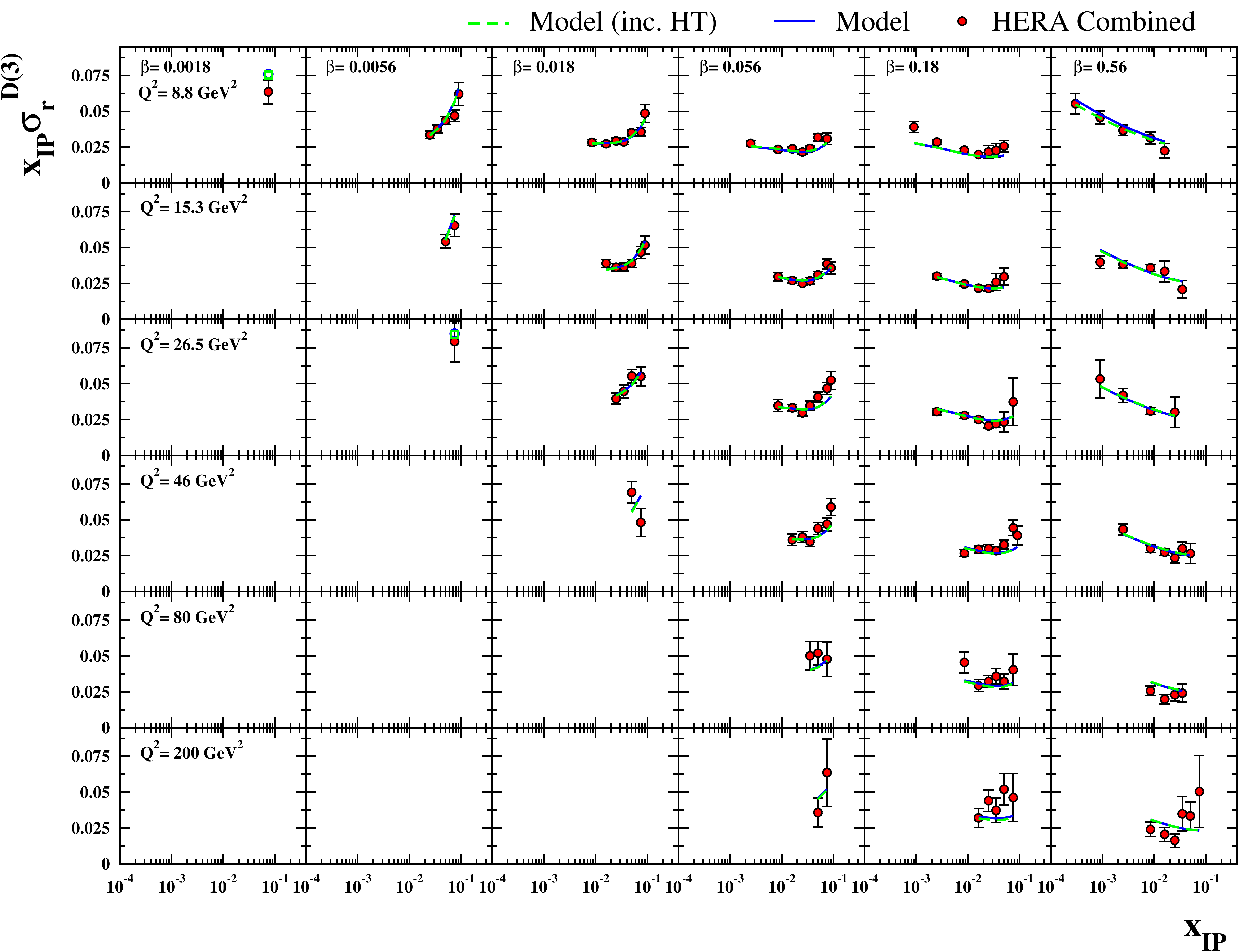}} 
	\begin{center}
		\caption{{\small Comparison between the experimental data on the diffractive reduced cross sections $x_{\pom} \sigma_r^{D(3)} (\beta, Q^{2}; x_{\pom})$ from the recent H1 and ZEUS combined dataset~\cite{Aaron:2012hua} and the corresponding NLO theoretical predictions from our NLO QCD fits without (solid curves) and with (dashed curves) considering HT effects. The dots show the central values of the experimental data points and the data errors are defined by adding in quadrature the systematical and statistical uncertainties.  } \label{fig:Combined-2012}}
	\end{center}
\end{figure*}
\begin{figure*}[htb]
	\vspace{0.50cm}
	\resizebox{0.60\textwidth}{!}{\includegraphics{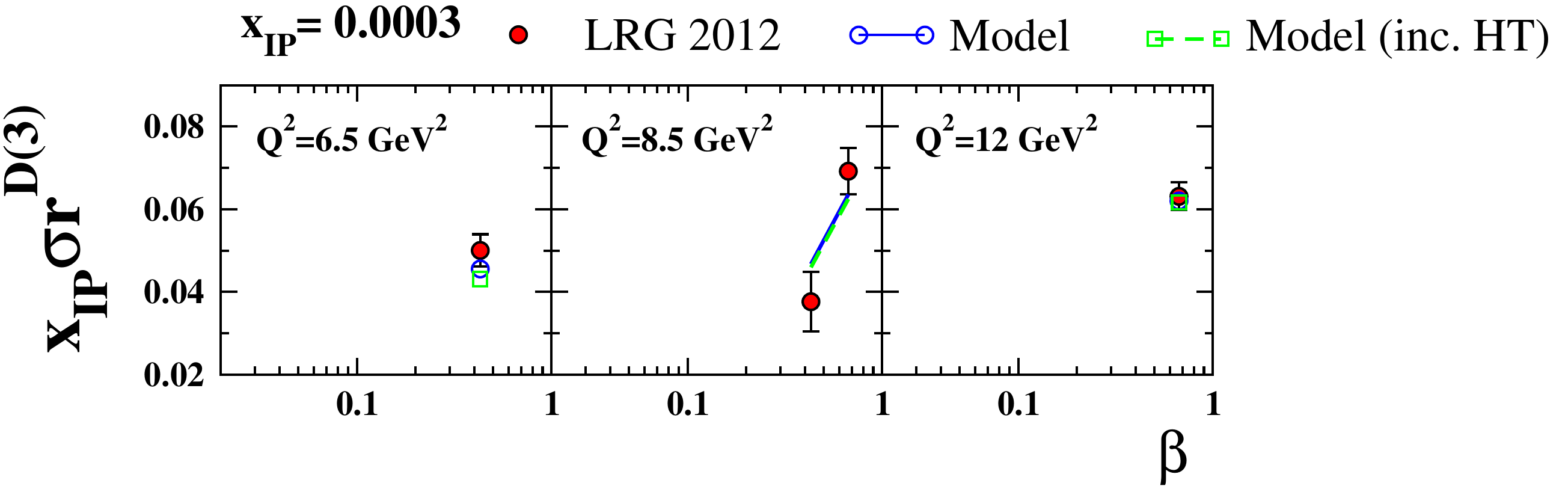}} 
	\begin{center}
		\caption{{\small  Comparison between the experimental data on the diffractive reduced cross sections $x_{\pom} \sigma_r^{D(3)} (\beta, Q^2; x_{\pom})$ from the H1-LRG-2012 datasets~\cite{Aaron:2012ad} and the corresponding NLO theoretical predictions from our NLO QCD fits without (solid curves) and with (dashed curves) considering HT effects. The comparisons have been done for the fixed value of $x_{\pom} = 0.0003$ and different photon virtuality $Q^2$. } \label{fig:LRG-2012-xp0003}}
	\end{center}
\end{figure*}
\begin{figure*}[htb]
	\vspace{0.50cm}
	\resizebox{0.60\textwidth}{!}{\includegraphics{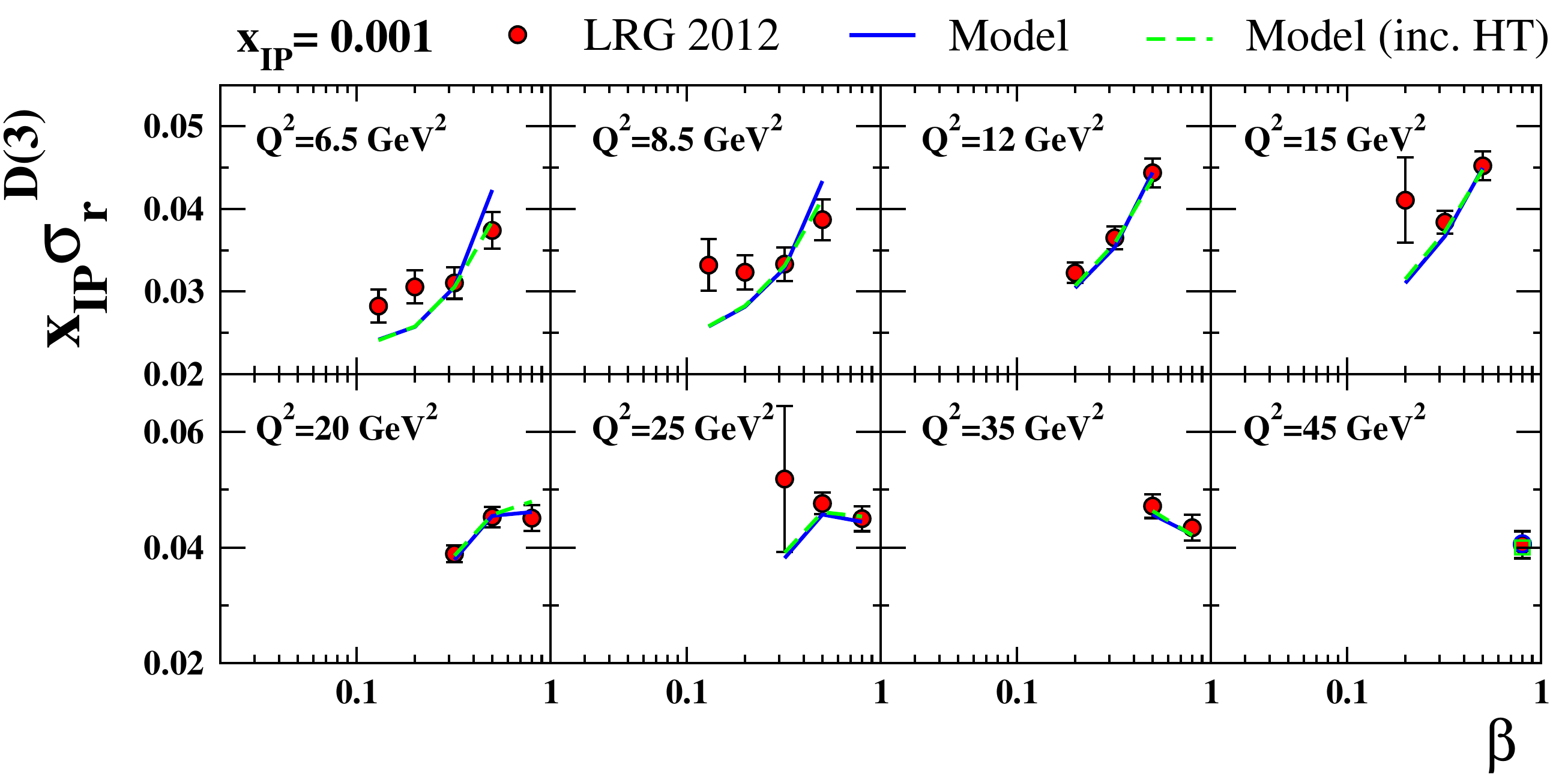}} 
	\begin{center}
		\caption{{\small Same as Fig.~\ref{fig:LRG-2012-xp0003} but for the fixed value of $x_{\pom} = 0.001$. } \label{fig:LRG-2012-xp001}}
	\end{center}
\end{figure*}
\begin{figure*}[htb]
	\vspace{0.50cm}
	\resizebox{0.60\textwidth}{!}{\includegraphics{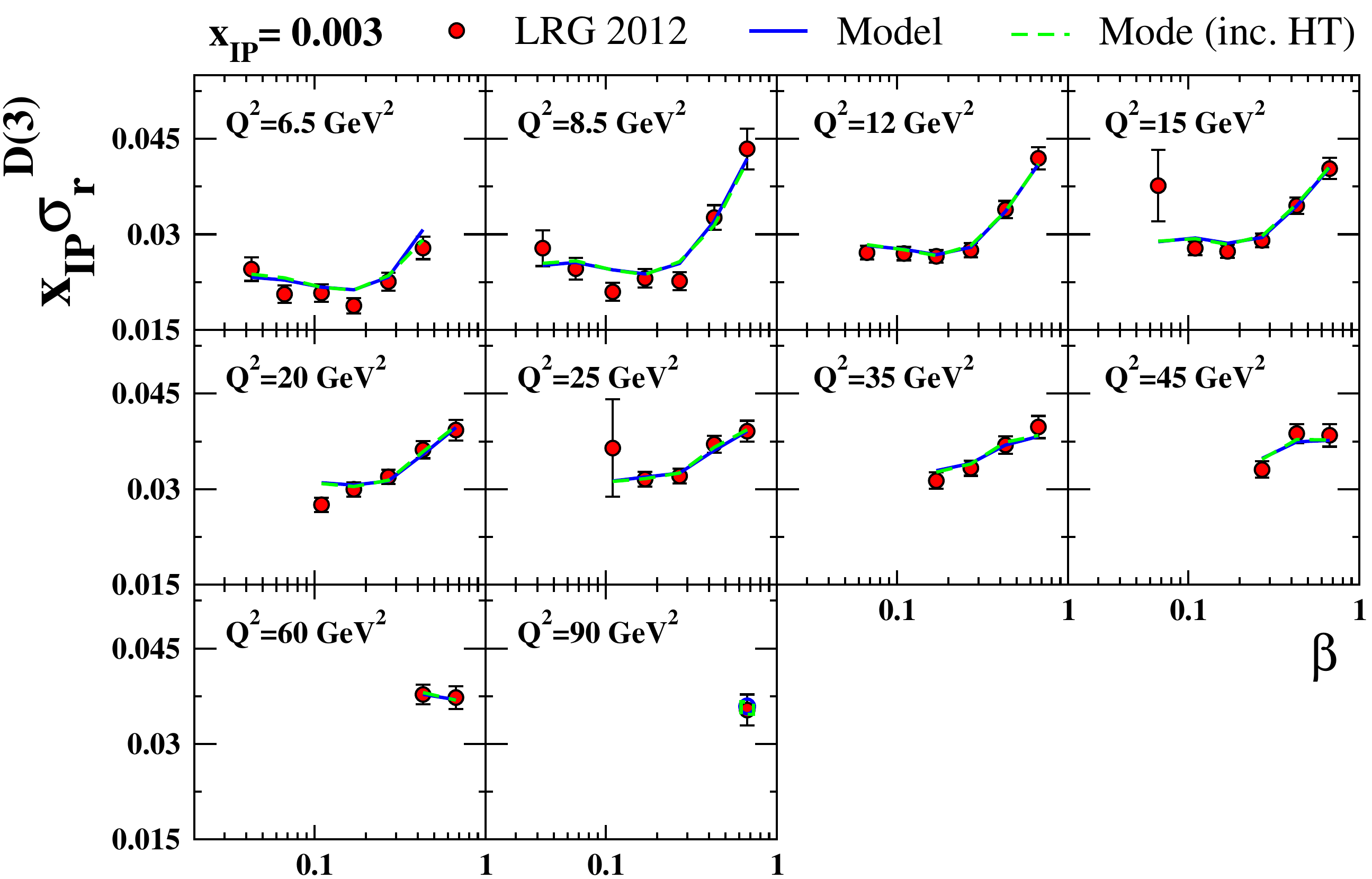}}  
	\begin{center}
		\caption{{\small  Same as Fig.~\ref{fig:LRG-2012-xp0003} but for the fixed value of $x_{\pom} = 0.003$. } \label{fig:LRG-2012-xp003}}
	\end{center}
\end{figure*}
\begin{figure*}[htb]
	\vspace{0.50cm}
	\resizebox{0.60\textwidth}{!}{\includegraphics{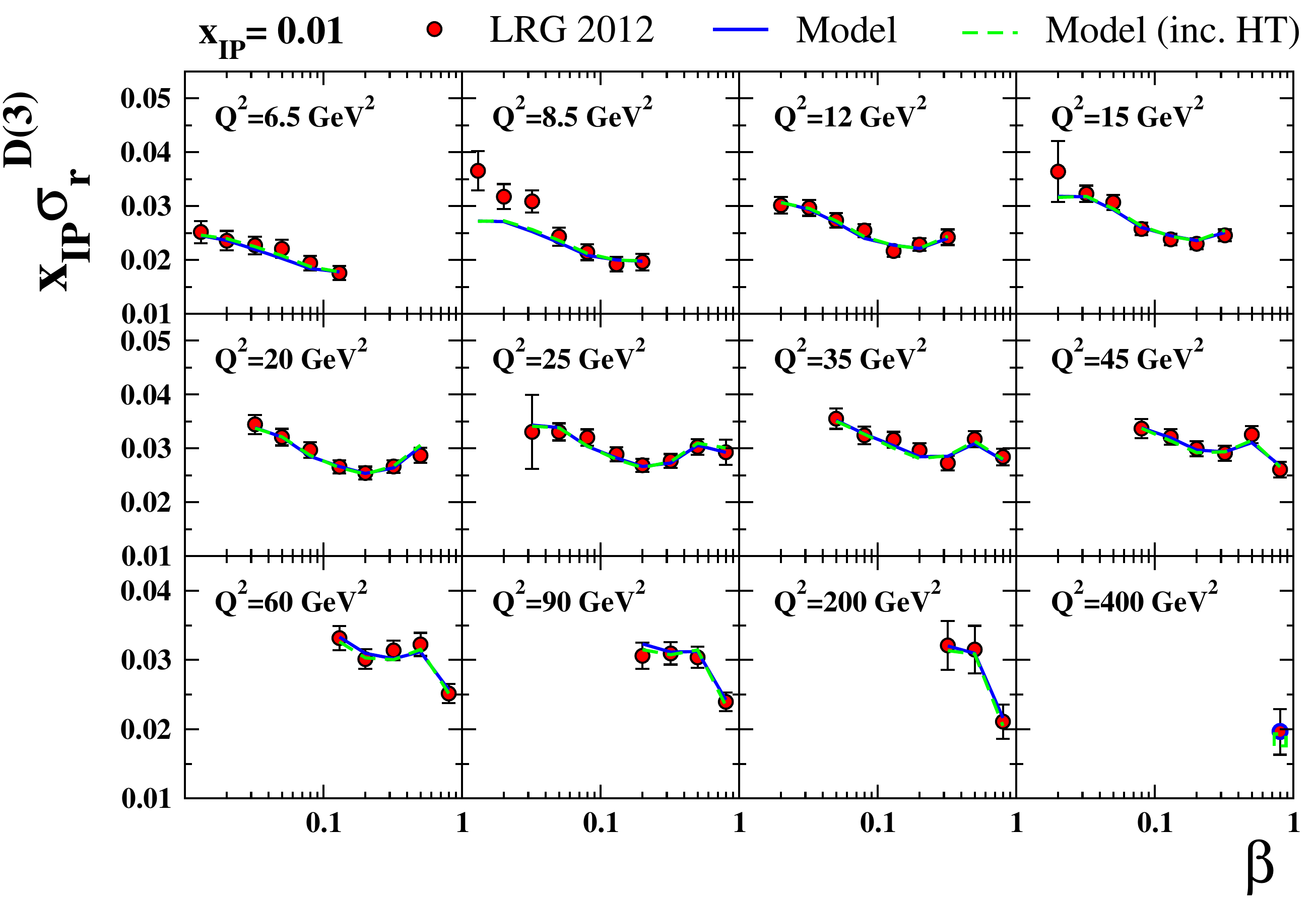}} 
	\begin{center}
		\caption{{\small  Same as Fig.~\ref{fig:LRG-2012-xp0003} but for the fixed value of $x_{\pom} = 0.01$. } \label{fig:LRG-2012-xp01}}
	\end{center}
\end{figure*}
\begin{figure*}[htb]
	\vspace{0.50cm}
	\resizebox{0.60\textwidth}{!}{\includegraphics{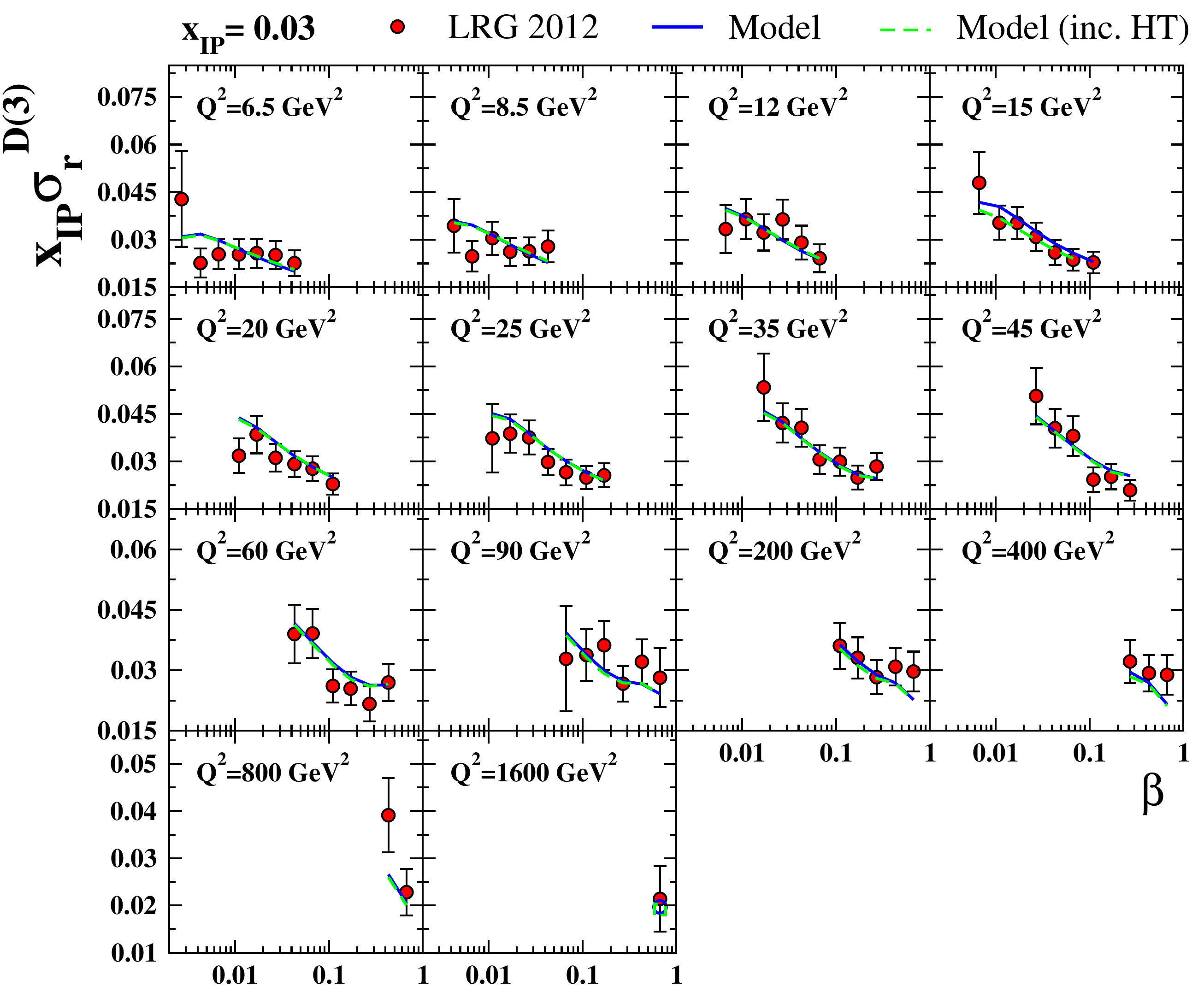}}
	\begin{center}
		\caption{{\small  Same as Fig.~\ref{fig:LRG-2012-xp0003} but for the fixed value of $x_{\pom} = 0.03$. } \label{fig:LRG-2012-xp03}}
	\end{center}
\end{figure*}

\begin{figure*}[htb]
	\vspace{0.20cm}
	\resizebox{0.480\textwidth}{!}{\includegraphics{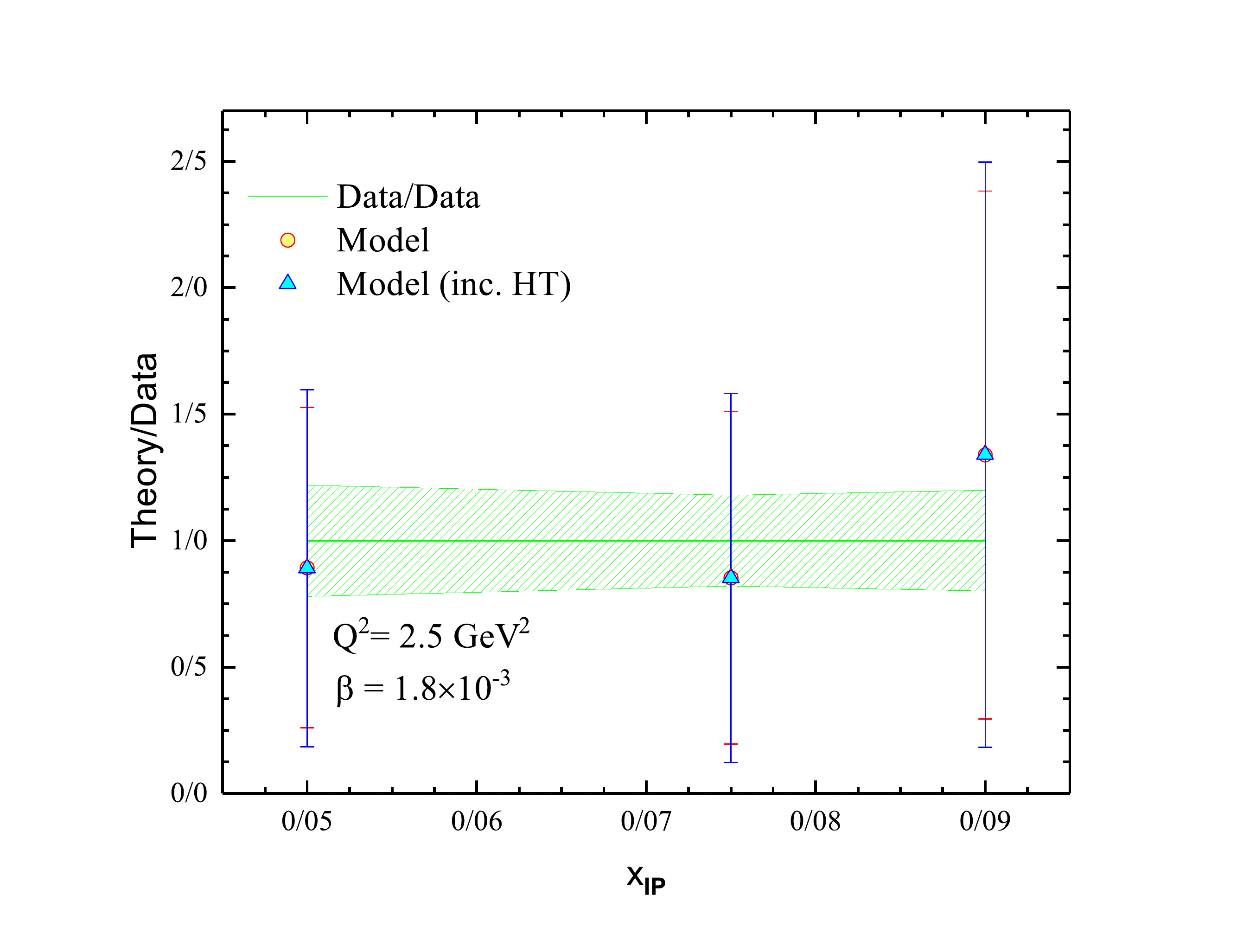}}  
	\resizebox{0.480\textwidth}{!}{\includegraphics{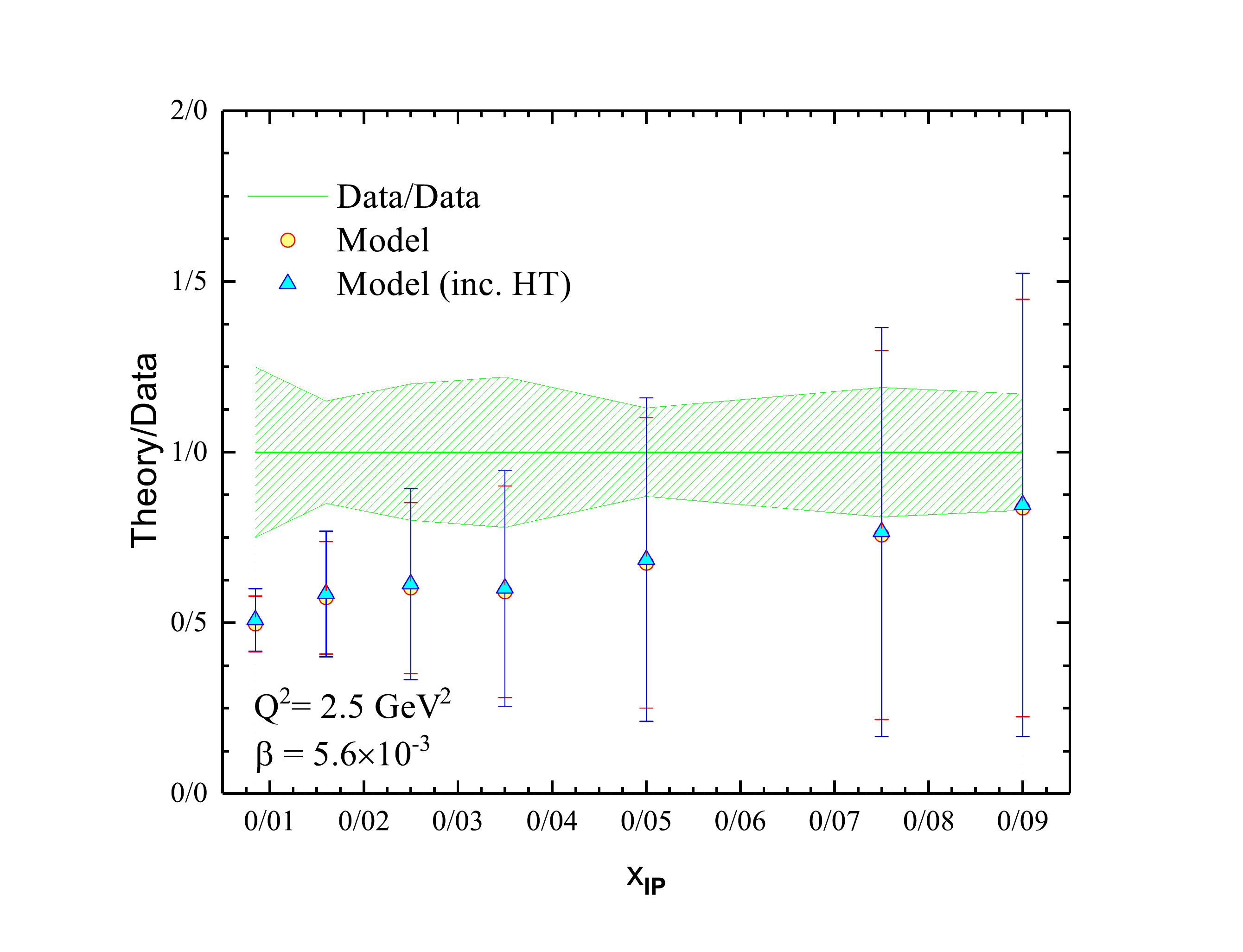}}  
	\resizebox{0.480\textwidth}{!}{\includegraphics{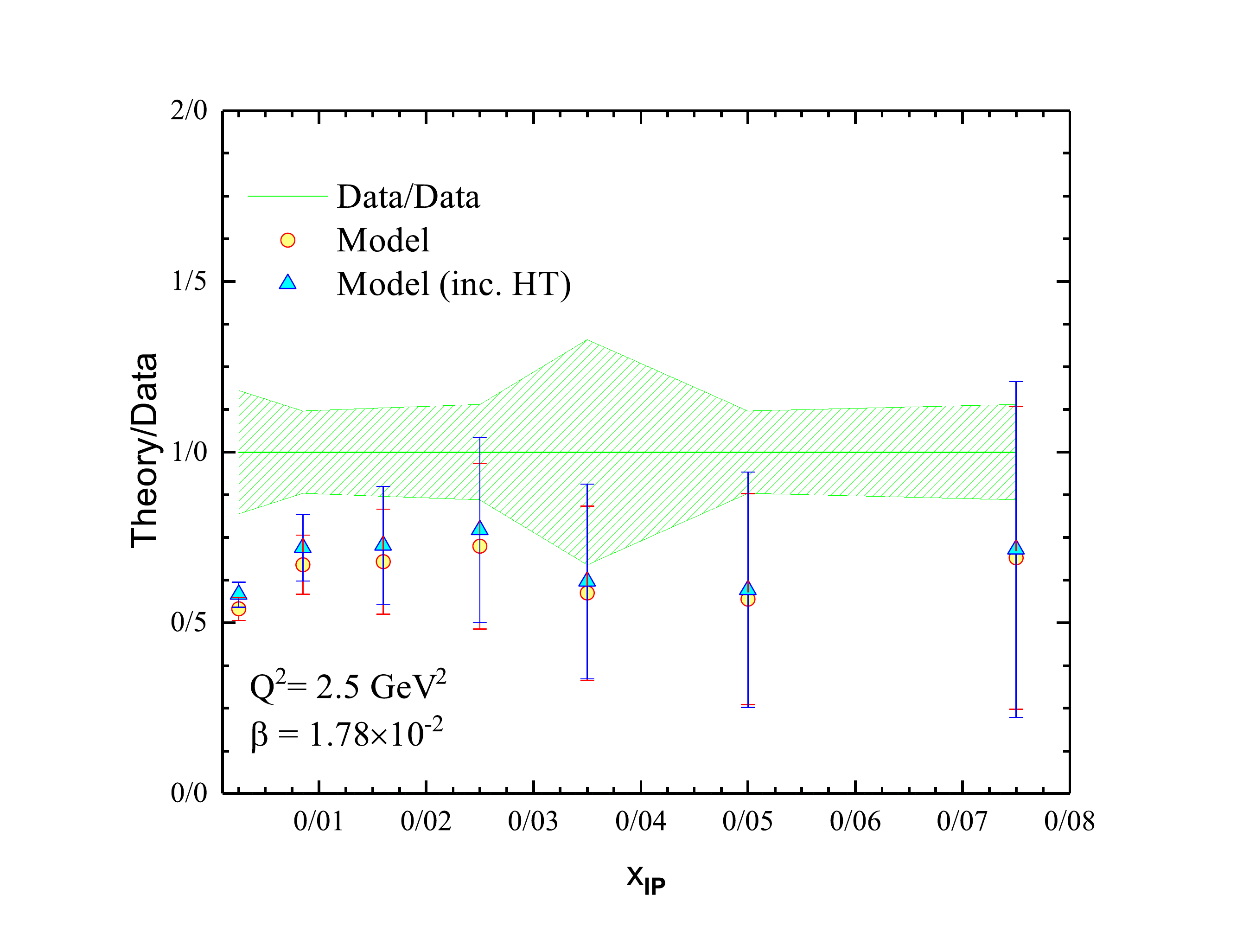}}  
	\resizebox{0.480\textwidth}{!}{\includegraphics{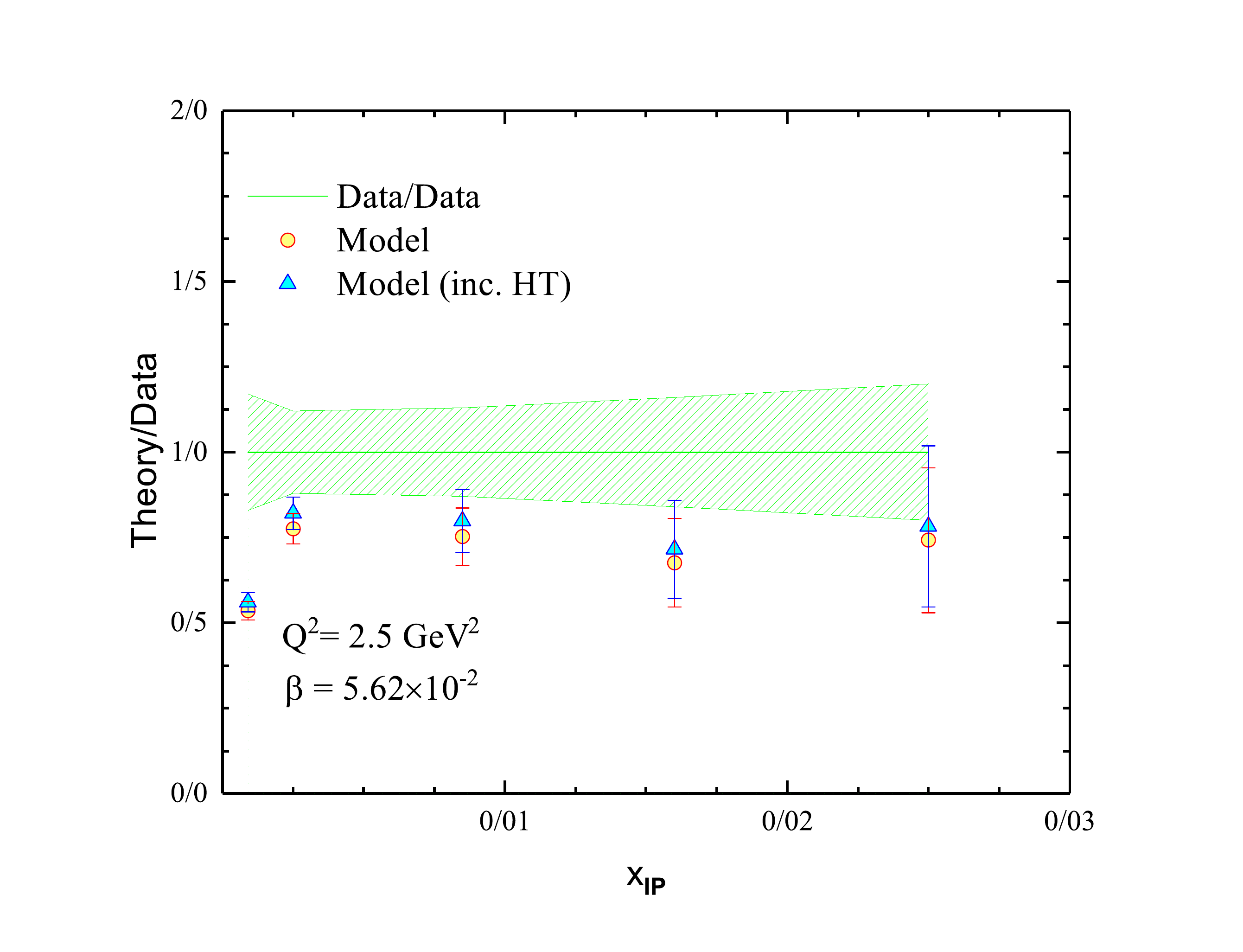}}  
	\resizebox{0.480\textwidth}{!}{\includegraphics{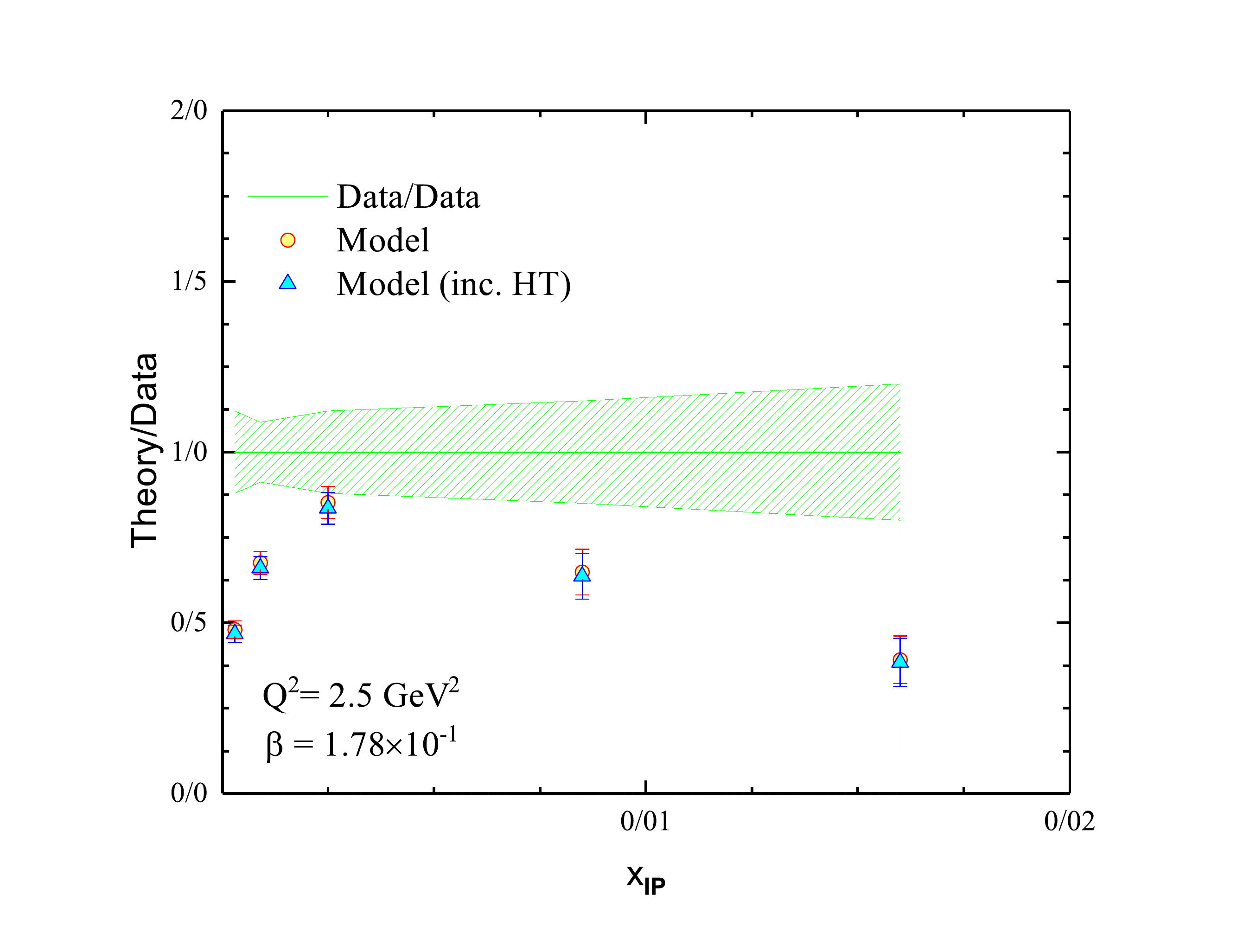}}  				
	\begin{center}
		\caption{{\small Comparison between the experimental data from the H1 and ZEUS combined dataset~\cite{Aaron:2012hua} at $Q^2$ = 2.5 GeV$^2$ and the corresponding theoretical predictions without (circle) and with (triangle) including HT effects as a theory to data ratios. The shaded bands correspond to the errors of experimental data points which are obtained  by adding systematical and statistical uncertainties in quadrature.} \label{fig:RatioQ2.5}}
	\end{center}
\end{figure*}
\begin{figure*}[htb]
	\vspace{0.20cm}
	\resizebox{0.480\textwidth}{!}{\includegraphics{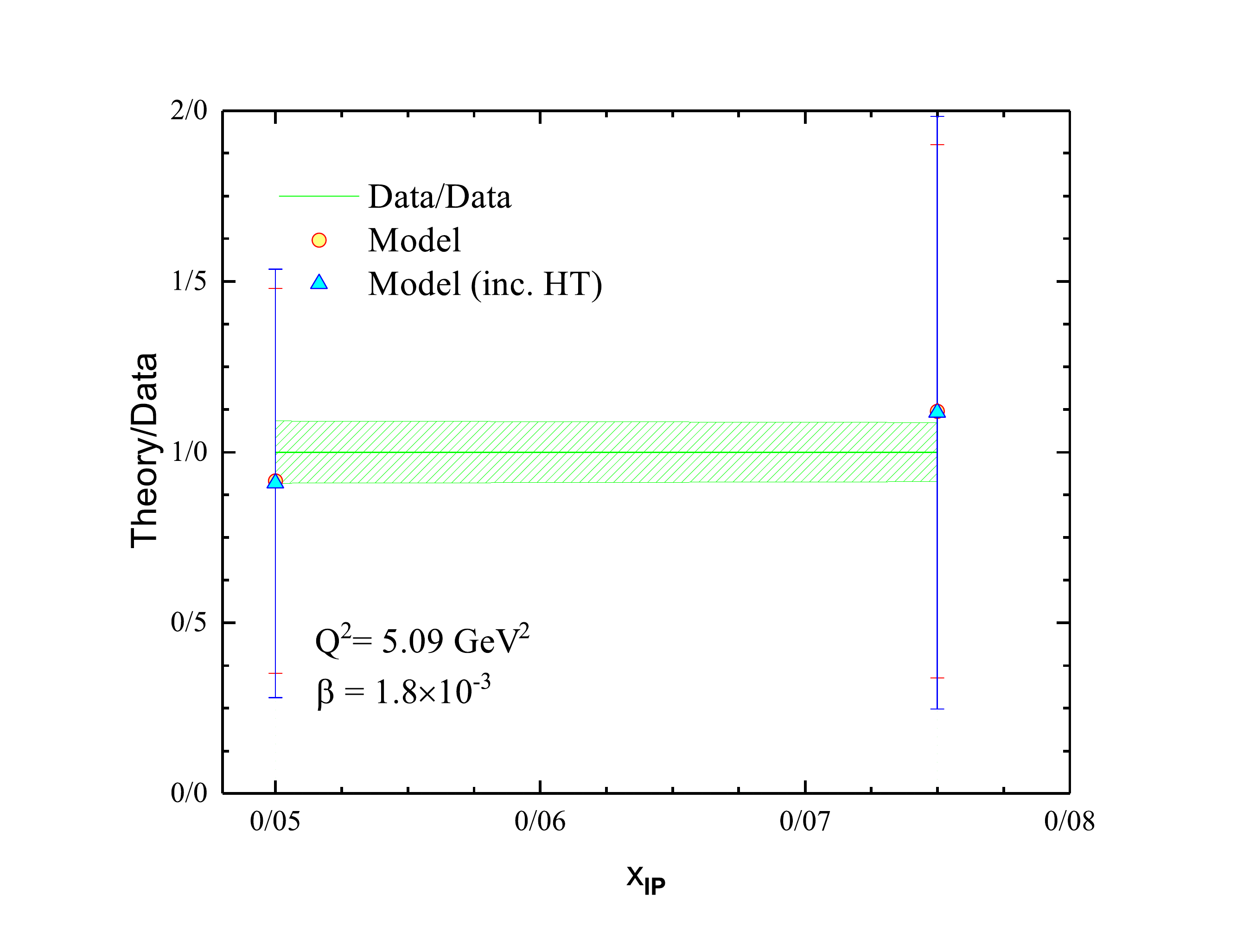}}  
	\resizebox{0.480\textwidth}{!}{\includegraphics{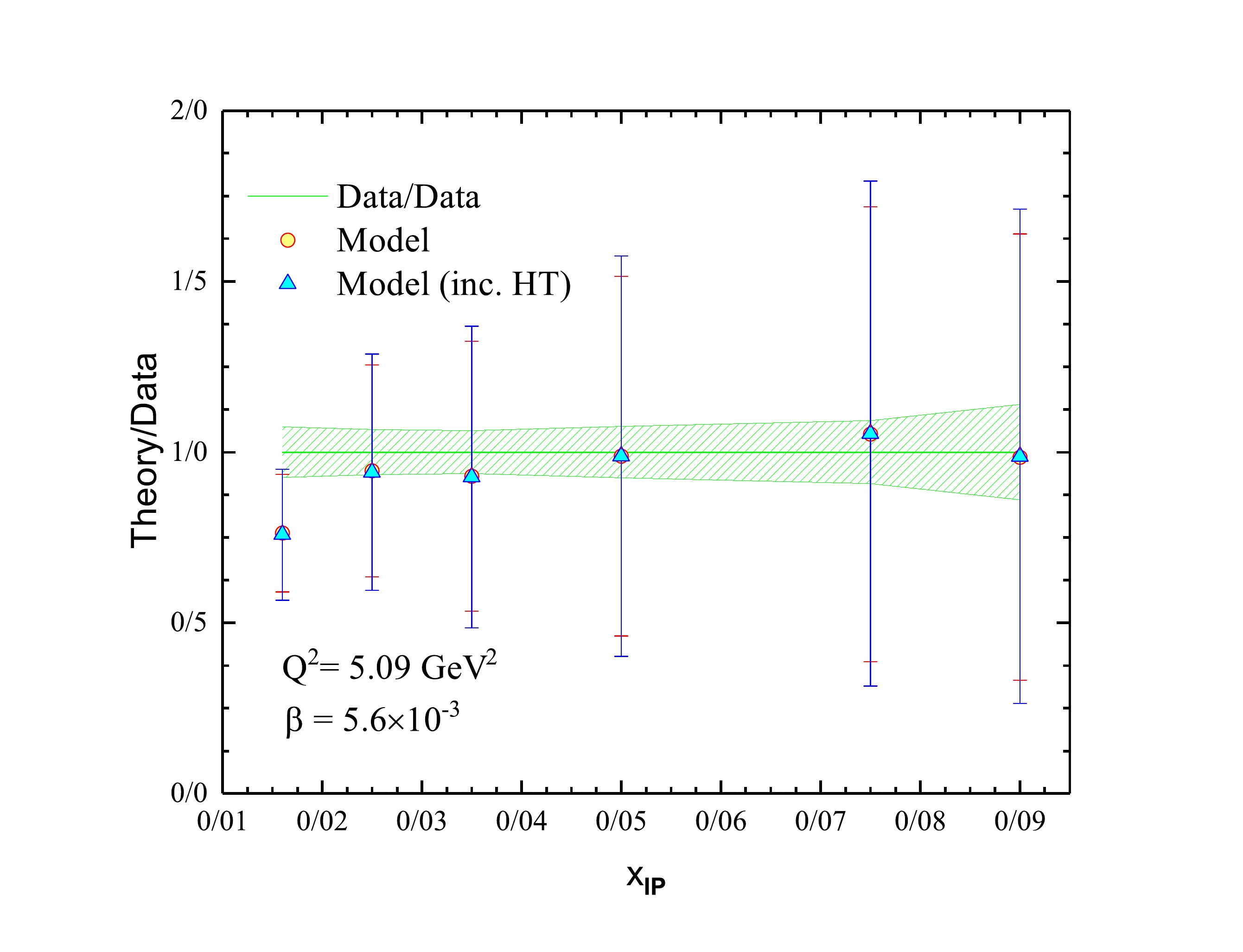}}  
	\resizebox{0.480\textwidth}{!}{\includegraphics{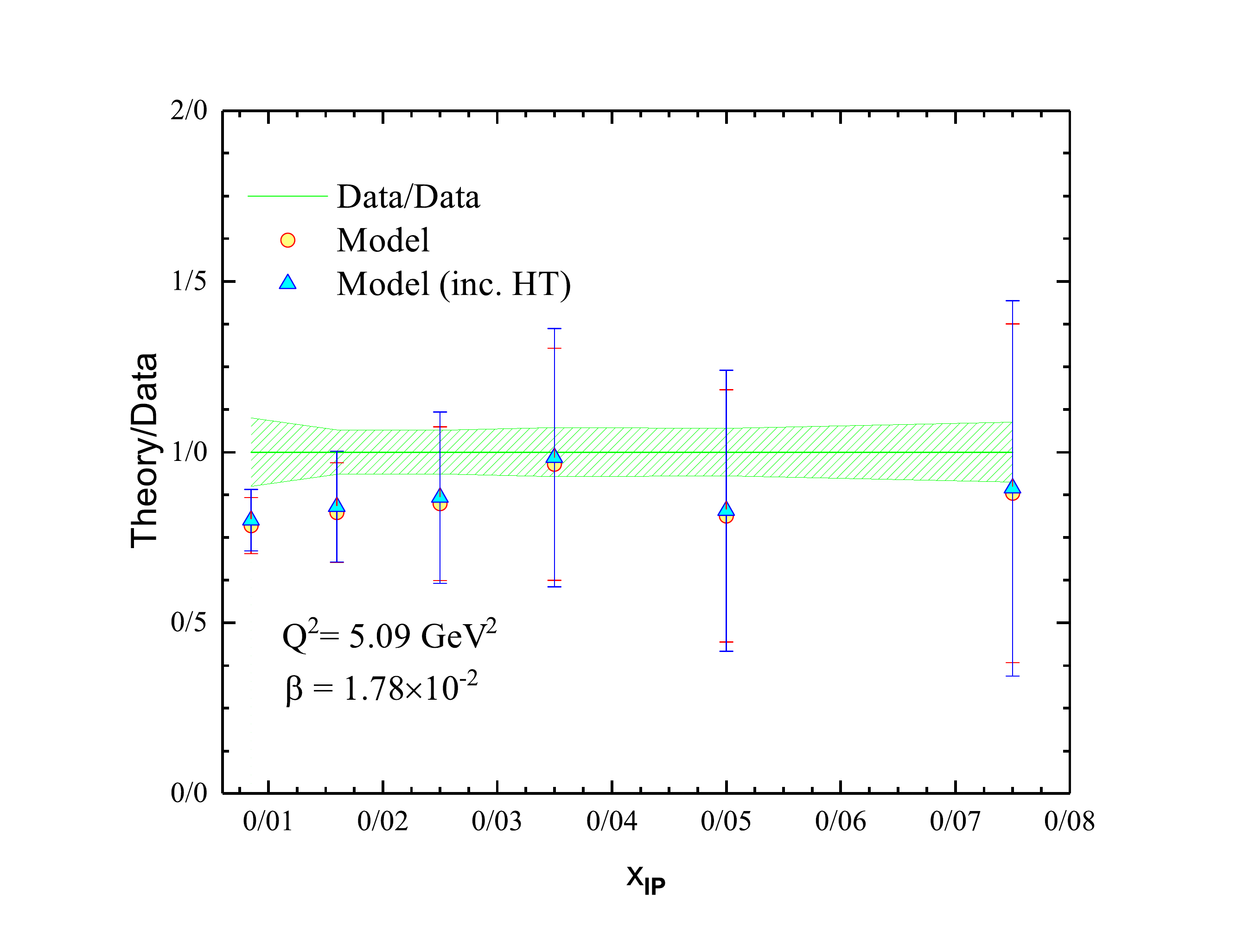}}  
	\resizebox{0.480\textwidth}{!}{\includegraphics{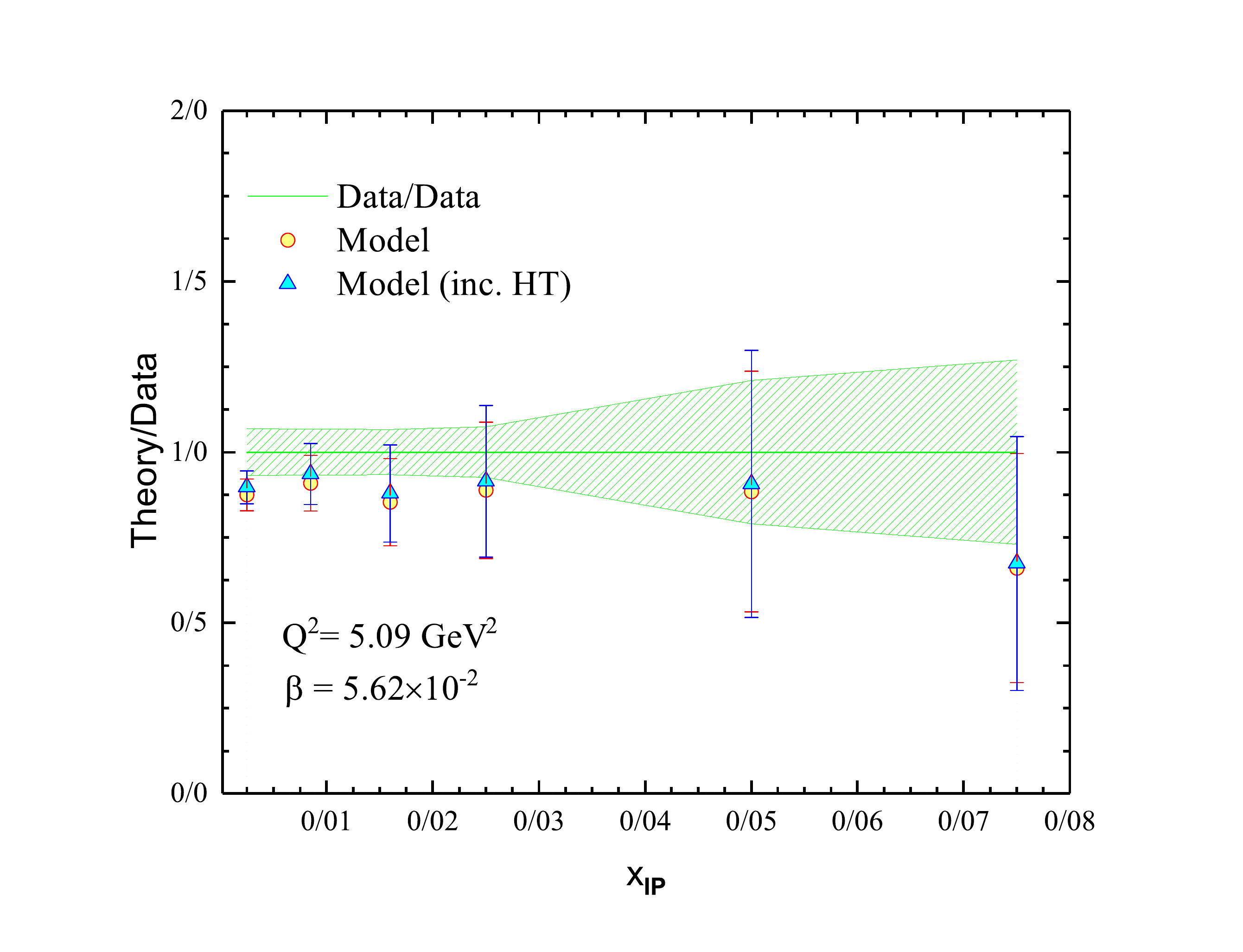}}  
	\resizebox{0.480\textwidth}{!}{\includegraphics{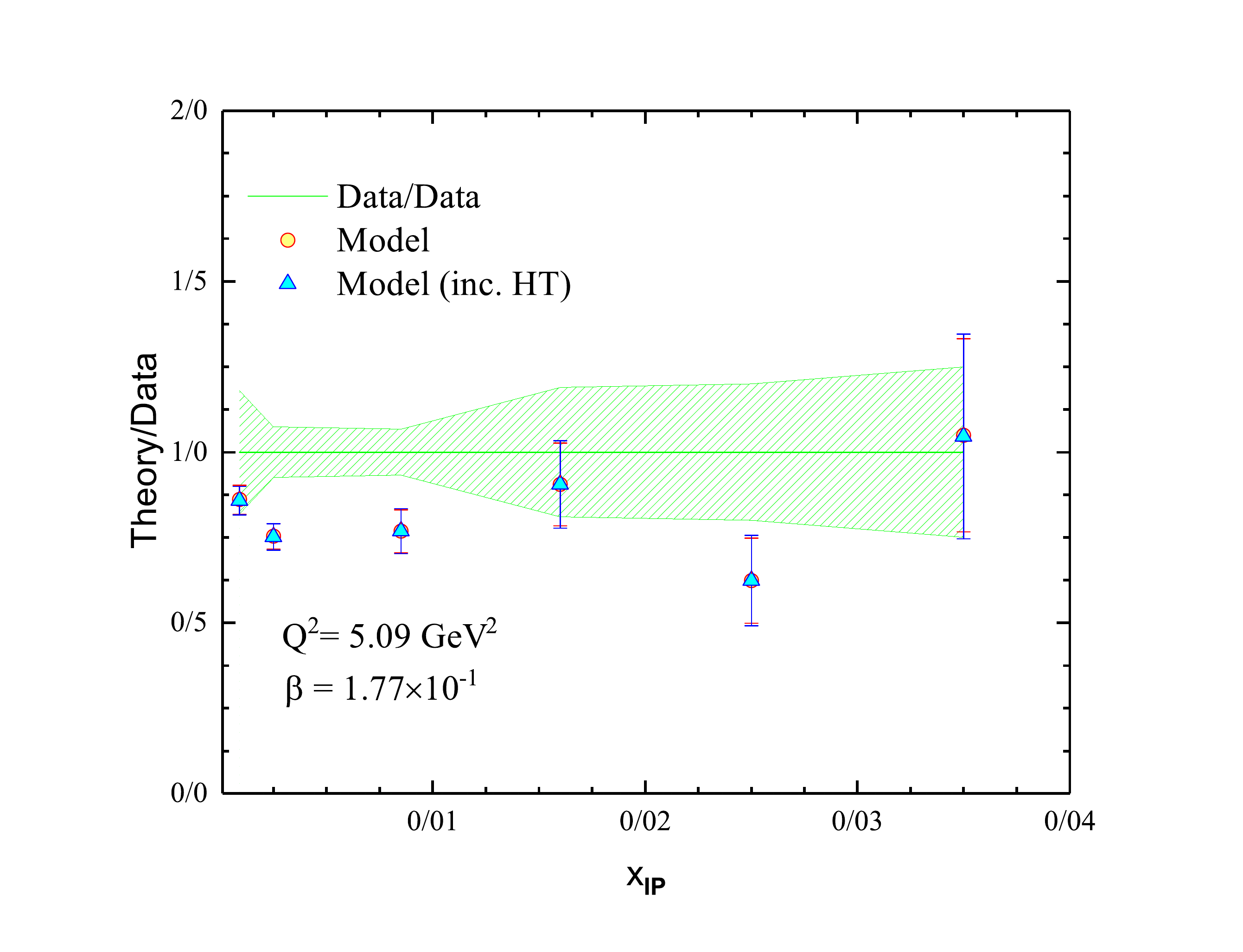}}  				
	\begin{center}
		\caption{{\small Same as Fig.~\ref{fig:RatioQ2.5} but for $Q^2$ = 5.09 GeV$^2$. } \label{fig:RatioQ5.09}}
	\end{center}
\end{figure*}
\begin{figure*}[htb]
	\vspace{0.20cm}
	\resizebox{0.480\textwidth}{!}{\includegraphics{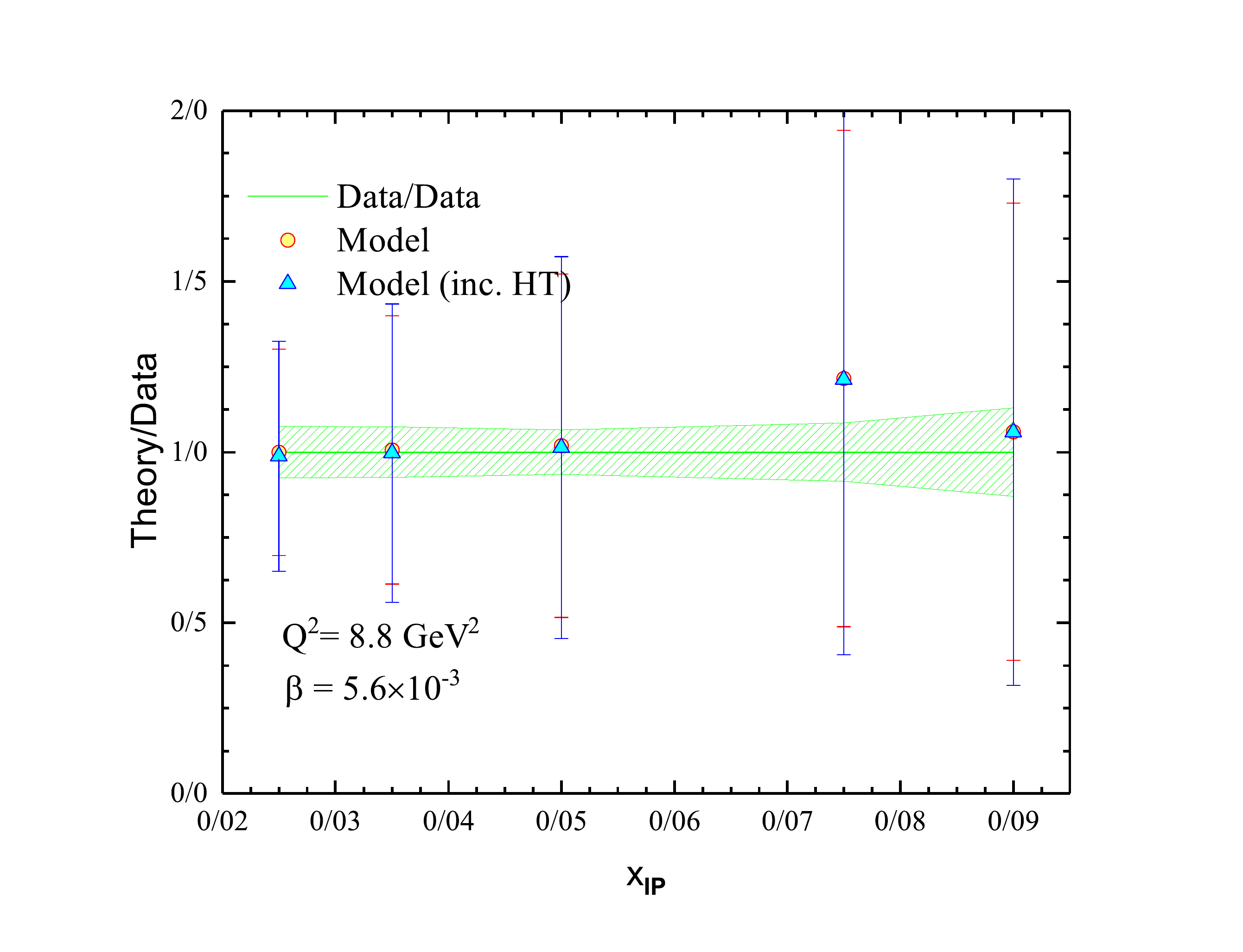}}  
	\resizebox{0.480\textwidth}{!}{\includegraphics{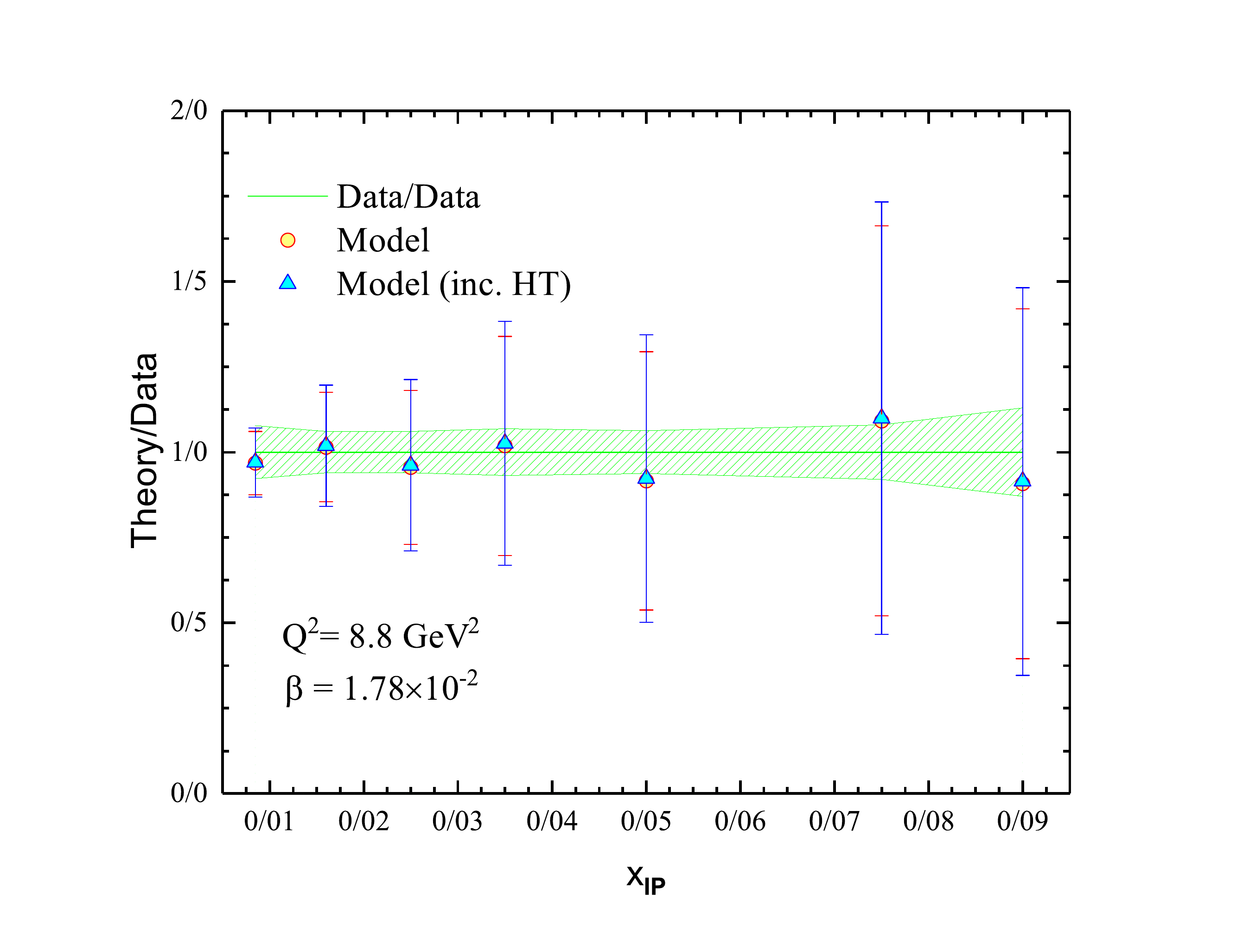}}  
	\resizebox{0.480\textwidth}{!}{\includegraphics{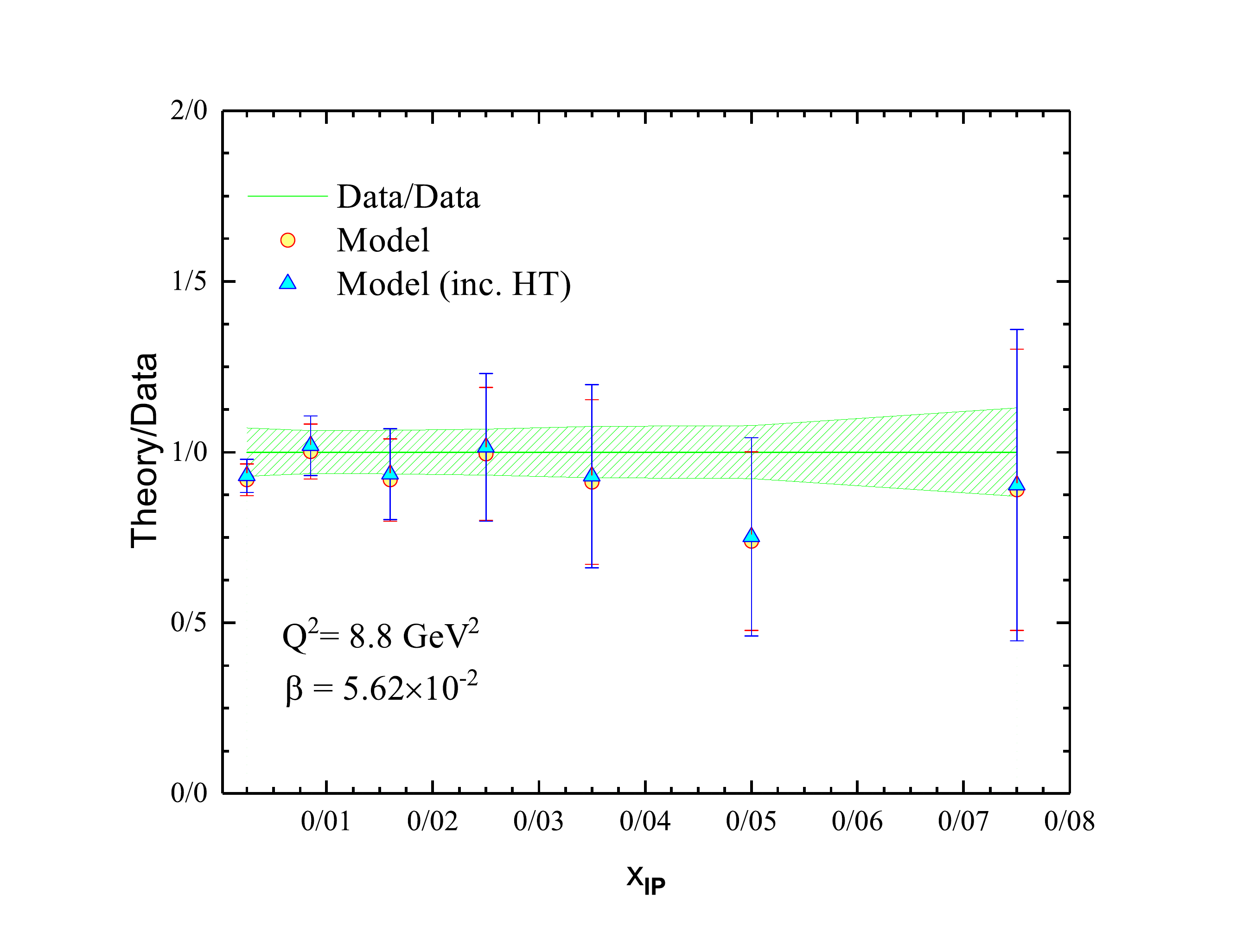}}  
	\resizebox{0.480\textwidth}{!}{\includegraphics{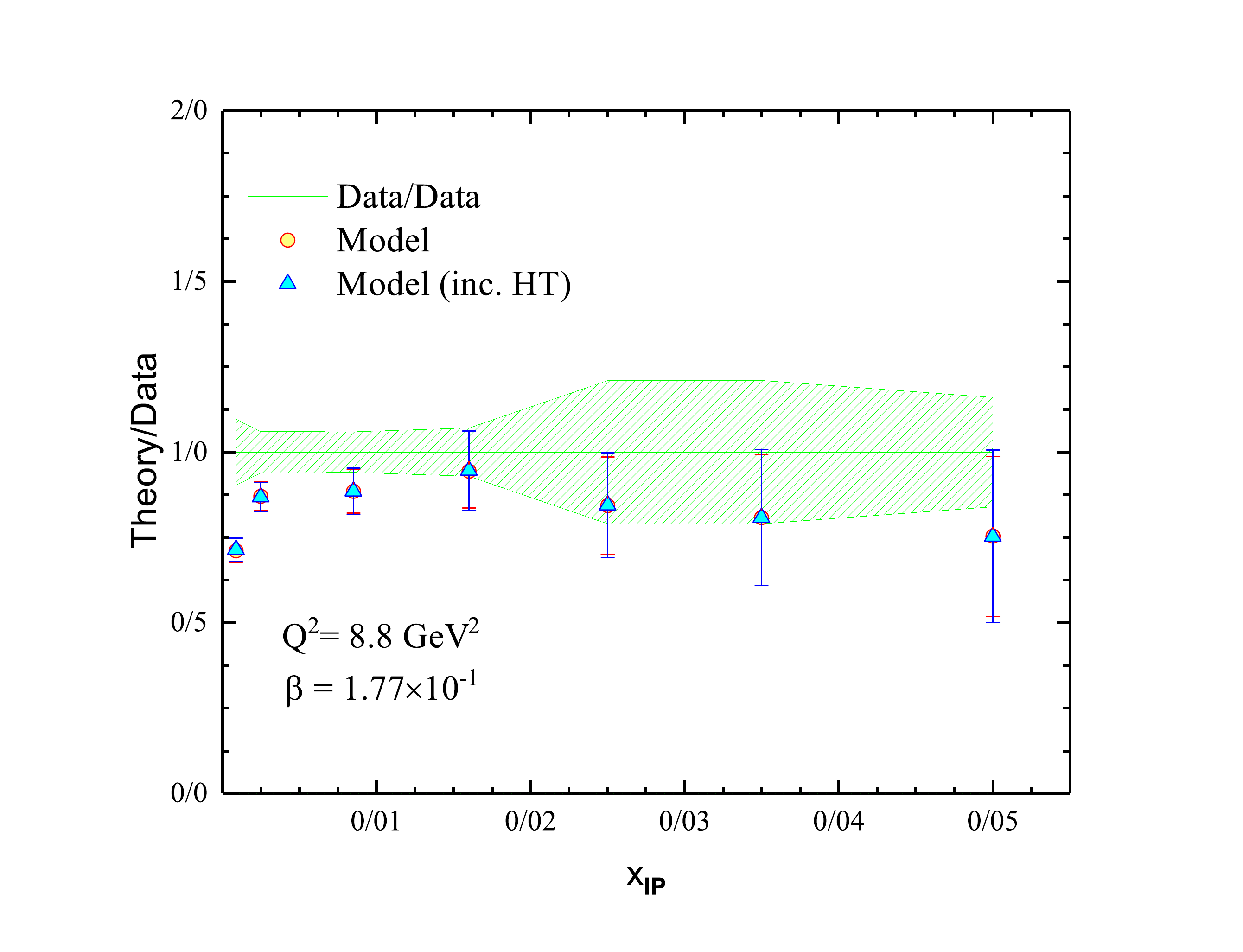}}  
	\resizebox{0.480\textwidth}{!}{\includegraphics{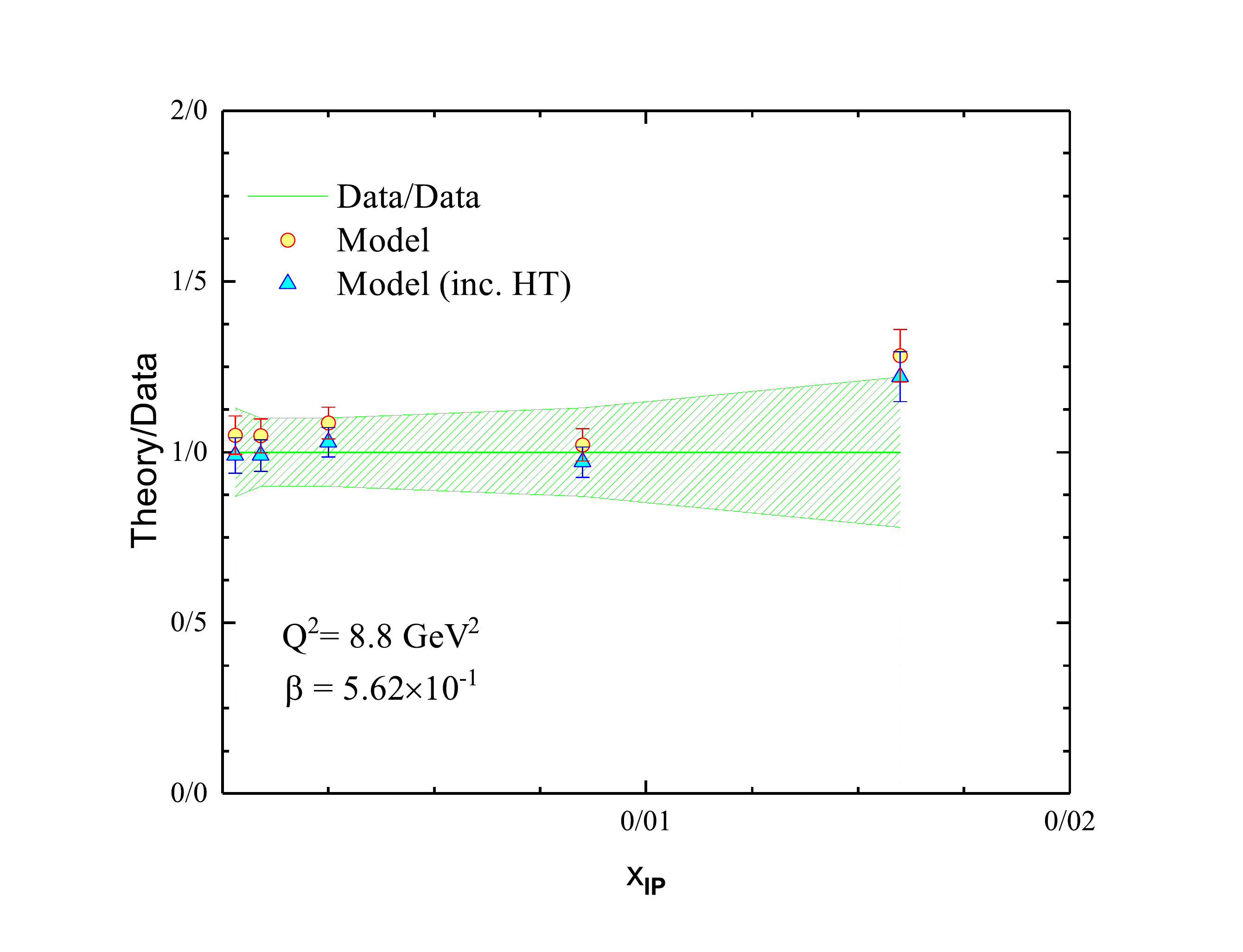}}  				
	\begin{center}
		\caption{{\small  Same as Fig.~\ref{fig:RatioQ2.5} but for $Q^2$ = 8.8 GeV$^2$. } \label{fig:RatioQ8.8}}
	\end{center}
\end{figure*}
\section{ Results of the diffractive PDFs analysis }\label{sec:Results}

After introducing the phenomenological framework and experimental diffractive DIS datasets used in the present analysis, in this section, we aim to present the results obtained for diffractive PDFs with and without considering the HT effects. In order to present more detailed discussions on our findings, we also compare the results with other available diffractive PDFs analyses from {\tt H1-2006}~\cite{Aktas:2006hy}, {\tt ZEUS-2010}~\cite{Chekanov:2009aa} and {\tt GKG18}~\cite{Goharipour:2018yov}. Several comparisons of the NLO theoretical predictions and experimental data will be presented to check the quality of the fits in different kinematics. We mainly focus the regions of phase space where the HT has the largest impact on the quality of data/theory comparison. 

We first start our discussions with the best fitted parameters obtained in this analysis. Table.~\ref{tab-dpdf-parameters}, contains the extracted fit parameters from our QCD analyses, {\tt Model} and  {\tt Model (inc. HT)}. Please note that the later contains the results obtained considering HT effects. As one can conclude from this table, almost all parameters well determined, except the $\alpha_g$ for the case of HT analysis. The HT parameters, $h_0$, $h_1$ and $h_2$, obtained in this fit along with the diffractive PDFs parameters also presented in the second part of the table. Parameters marked with the (*) have been fixed at their best fitted values. In order to give more flexibility to the parameterizations of Eq.~\eqref{eq:fD}, we let all parameters to be free in the fit. Our investigations and the obtained results indicate that currently available diffractive DIS datasets would not be able to put further constraints on the $z$ dependence of the diffractive PDFs, suggesting more accurate experimental measurements need to be taken into account. In any case, we see from the first step of the minimization procedure that one can not determine the $\eta$ parameter well enough. Hence, we prefer to fix this parameter to zero both for the quark and gluon densities.

%
%
\begin{table}[ht!]
\begin{center}
\caption{\small Parameters obtained with the different fits at the initial scale $Q_0^2 = 1.8 \, \gev^{2}$ along with their experimental uncertainties. Parameters marked with the (*) are fixed at their best fitted values. }
\begin{tabular}{ c | c | c }
\hline \hline
Parameters	  & {\tt Model}       & {\tt Model (inc. HT)}        \\  \hline \hline
$\alpha_q$    & $0.376 \pm 0.029$ & $0.436 \pm 0.032$  \\ 
$\beta_q$     & $1.699 \pm 0.080$ & $1.541 \pm 0.072$  \\ 
$\gamma_q$    & $0.607 \pm 0.042$ & $0.723 \pm 0.042$  \\ 
$\eta_q$      & $0.0^*$           & $0.0^*$            \\   

$\alpha_g$    & $2.166 \pm 0.468$ & $6.244 \pm 2.282$  \\ 
$\beta_g$     & $0.545 \pm 0.095$ & $0.831 \pm 0.137$  \\ 
$\gamma_g$    & $0.741 \pm 0.215$ & $2.400 \pm 0.509$   \\ 
$\eta_g$      & $0.0^*$           & $0.0^*$            \\   

$\alpha_{\pom}(0)$  & $1.091 \pm 0.003$  & $1.090 \pm 0.0031$    \\
$\alpha_{\reg}(0)$  & $0.436 \pm 0.074$  & $0.437 \pm 0.080$      \\
$A_{\reg}$          & $6.279 \pm 2.387$  & $6.279 \pm 2.582$       \\  \hline \hline

$h_0$       & $0.0$  & $-25.798$    \\
$h_1$       & $0.0$  & $2.536$      \\
$h_2$       & $0.0$  & $-1.381$       \\  \hline \hline

$\alpha_s(M_Z^2)$   & $0.1176^*$           & $0.1176^*$             \\
$m_c$               & $1.40^*$             & $1.40^*$               \\
$m_b$               & $4.750^*$            & $4.75^*$               \\ 	\hline \hline
\end{tabular}
\label{tab-dpdf-parameters}
\end{center}
\end{table}
%
%

In order to have a better insight into which datasets led to better fit quality, let us now discuss our results in term of individual $\chi^2$ we obtained for each experiments and for both of our analyses.  
Table~\ref{tab:chi2} contains the results of $\chi^2$ for both analyses of diffracive DIS data performed in the present work, with and without considering the HT terms. For each dataset, we have presented the related reference and also the value of $\chi^2$ divided to the number of analyzed data points ($ \chi^2 /N_{\textrm{pts}}$). Note that the last row of the table contains the values of $\chi^2$ per number of degrees of freedom, $\chi^2/\textrm{dof}$, for both analyses. As can be seen from this table, a reasonable improvement in overall fit quality has been obtained after considering the HT contributions in the analysis. One can also conclude that the main improvements in the fit quality come mainly from the {\tt H1-LRG-12}~\cite{Aaron:2012ad} experimental data. For this dataset, we also see a much better $\chi^2$ after including the HT corrections.
The improvement due to the HT terms also visible for the high precision {\tt H1/ZEUS combined}~\cite{Aaron:2012hua} experimental data.

\begin{table*}[ht!]
\caption{ \small The values of $\chi^2/N_{\text{pts}}$ for the datasets included in the analysis of diffractive PDFs without considering HT (second column) and by considering this term in the fit (third column). } \label{tab:chi2}
\begin{tabular}{l c c c }
\hline \hline
& {\tt Without HT}             & {\tt With HT}  \\ \hline
Experiment & $\chi^2/N_{\text{pts}}$  & $\chi^2/N_{\text{pts}}$ 
\tabularnewline
\hline
{\tt  H1-LRG-11} $\sqrt{s} = 225$~GeV~\cite{Aaron:2012zz} & 15.595/20 &  16.421/20  \\	
{\tt  H1-LRG-11} $\sqrt{s} = 252$~GeV~\cite{Aaron:2012zz} & 16.934/19 &  17.834/19  \\		
{\tt  H1-LRG-11} $\sqrt{s} = 319$~GeV~\cite{Aaron:2012zz} & 7.220/12 &  6.646/12  \\			
{\tt H1-LRG-12}~\cite{Aaron:2012ad}        & 179.323/267      &  165.654/267  \\	
{\tt  H1/ZEUS combined}~\cite{Aaron:2012hua} & 168.553/181  &  163.606/181  \\	 \hline
\multicolumn{1}{c}{~\textbf{$\chi^2/{\text {dof}}$}~}       &    $~390.46/371=1.052~$ &    $~376.079/371=1.013~$  \\  \hline\hline
\end{tabular}
\end{table*}
%
%

After reviewing the individual $\chi^2$ for each dataset to judge the overall fit quality, we now discuss the extracted diffractive PDFs for both analyses. As we mentioned, one of the key ingredients in the strategy pursued in the this analysis is the inclusion of HT terms. Hence, it should be interesting to see the significant change in the shape of the distributions after including the HT corrections. 

In Fig.~\ref{fig:DPDF-Q0}, our diffractive PDFs have been shown as a function of momentum fraction $z$ obtained at the input scale of $Q_0^2 = 1.8 \, \gev^2$ for both of our analyses with and without the HT corrections. The error bands represent the uncertainty estimation coming from the experimental errors. It is also worth noticing here that for calculating of uncertainty bands for our diffractive PDFs and all theoretical predictions we use the standard ``Hessian'' error propagation~\cite{Martin:2003sk,Pumplin:2001ct,Martin:2009iq,deFlorian:2011fp,Schmidt:2018hvu,Eskola:2009uj} with $\Delta\chi^2 = 1$. For the case of predictions including HT effects, the related theoretical uncertainties are also included. As one can see in this plot, the inclusion of HT term, significantly change the shape of gluon density. One can see an enhancement for the small value of $z$ and reduction for the larger value of $z$. For the quark density, one can also see an enhancement for the small value of $z$. It is also worthwhile mentioned here that the kinematic region where the impact of the HT correction is most important is precisely the high region of $z$ at low $Q^2$, which mostly affect the shape of the gluon density. In term of comparison of error bands, one can see that the uncertainty bands for both of our analyses are almost similar in size, though a rather significant reduction is seen for gluon density at large values of $z$ after including HT effects.

At this point, it is also enlightening to compare the diffractive PDFs obtained in this study with those of other available groups. These comparisons are shown in Fig.~\ref{fig:DPDF-Q6-20-200} with the results from  {\tt H1-2006 Fit B}~\cite{Aktas:2006hy} and {\tt ZEUS-2010 Fit SJ}~\cite{Chekanov:2009aa} for gluon and quark densities in three photon virtuality of $Q^2 = 6\,, 20$ and $200$ GeV$^2$. The H1 analysis has been obtained with the H1 LRG data while ZEUS used for the first time the diffractive dijet production datasets. The error bands represent our uncertainty estimation using the ``Hessian'' method.
The comparison of the results reveals the following conclusions: For the case of diffractive quark PDFs, one see the same patterns for all results. They show a pick around the medium value of $z$, $z \sim 0.6$. However our results with the HT corrections are enhanced compared to those of other results for the region of $z < 0.6$. For the gluon density, the comparisons have shown in the left panels of Fig.~\ref{fig:DPDF-Q6-20-200} for $Q^2 = 6\,, 20$ and $200$ GeV$^2$. The comparisons between the gluon densities show very different patterns. The different shapes for the gluon density reflect some reasons such as the different datasets used in these analyses. We have used the most recent H1 and ZEUS combined dataset while {\tt H1-2006} used some older H1 LRG dataset. It is also worth mentioning the impact of the diffractive dijet production data 
on the gluon density function which used by the {\tt ZEUS-2010} QCD analysis. The diffractive dijet production should also lead to the reduction of the gluon uncertainty. As one can expect From Fig.~\ref{fig:DPDF-Q6-20-200}, for higher value of photon virtuality $Q^2 = 200$ GeV$^2$, all distributions follow the same trends. 

It would be also interesting to compare the extracted quark and gluon densities with the most recent {\tt GKG18}~\cite{Goharipour:2018yov} diffractive PDFs analysis. In Fig.~\ref{fig:GKG}, the comparison is shown for quark and gluon densities as a function of $z$ and for Q$^2$ = 6 GeV$^2$.  Like for the case of Fig.~\ref{fig:DPDF-Q6-20-200}, we see the same trend here and hence the same conclusions hold for the comparisons between our results and {\tt GKG18} analysis. {\tt GKG18} used the same datasets in their analysis with different and bigger kinematical cut on $Q^2_{\rm min}$, and hence the number of data points that we use in present analysis is more than {\tt GKG18} analysis. One of the main important results can be concluded from this figure is the significant reduction in gluon density at medium and large $ z $ regions due to including HT effects in the analysis.

Although the main results of this analysis have been presented up to now, there are still other important discussions remain in term of the data/theory comparisons. To start our discussions, in the following we present a detailed comparisons of our NLO theory predictions with almost all analyzed datasets in other to judge the fit quality, and then, to see the effect arising from the HT correction in some certain kinematical regions which are sensitive to this term. To this end, in Fig.~\ref{fig:Combined-2012} we compare our NLO theory prediction for the diffractive reduced cross sections $x_{\pom} \sigma_r^{D(3)} (\beta, Q^2; x_{\pom})$ with the recent H1 and ZEUS combined datasets~\cite{Aaron:2012hua}. In Figs.~\ref{fig:LRG-2012-xp0003}, \ref{fig:LRG-2012-xp001}, \ref{fig:LRG-2012-xp003}, \ref{fig:LRG-2012-xp01} and \ref{fig:LRG-2012-xp03} the NLO theory predictions have been also shown as a function of $\beta$ for some selected values of $Q^{2}$ and for four representative bins of $x_{\pom} = 0.01\,, 0.03\,, 0.001\,$ and $0.003$. The dots show the central values of the experimental data points and the data errors are defined by adding in quadrature the systematical and statistical uncertainties. These comparisons have been done for a wide range of $\beta$, $x_{\pom}$ and $Q^2$ for both of our analyses with and without the presence of HT correction. At the level of individual datasets and as expected by the $\chi^2$ values listed in Table.~\ref{tab:chi2}, these plots clearly show that the quality of the description between our NLO theory predictions and all the H1/ZEUS Combined and H1-LRG-2012 datasets analyzed in this study are quite acceptable. 

With the agreements between data and theory established up to now, we are in a position to quantify in more details the effect of HT corrections on the data/theory comparisons we describe in more detail below. However, from Table.~\ref{tab:chi2} one expects a better fit quality after inclusion of HT terms. From the data versus theory comparisons, deviation between our results with and without the HT corrections for some certain region of $Q^{2}$ and $\beta$ can clearly be observed, for example at $Q^2$ = 6.5 and 8.5 GeV$^2$ in Fig.~\ref{fig:LRG-2012-xp001} or  at $Q^{2}$ = 15 GeV$^2$ in Fig.~\ref{fig:LRG-2012-xp03}. Let us now present comparisons between the experimental data and the corresponding theoretical predictions as a theory to data ratios. In Figs.~\ref{fig:RatioQ2.5},~\ref{fig:RatioQ5.09} and~\ref{fig:RatioQ8.8}, we have shown these ratio in the presence of both of our results for some selected value of $Q^2$ = 2.5, 5.09 and 8.8 GeV$^2$, respectively. The shaded bands correspond to the errors of experimental data points which are obtained by adding the systematical and statistical uncertainties in quadrature. Here, circle and triangle stand for the corresponding theoretical predictions for  our analyses without and with the HT correction, respectively. Here the lack of agreements between theory and data can be traced to the low-$Q^{2}$ region of the diffractive reduced cross section ratios. The disagreement at low-$Q^{2}$ region is due to that we imposed a $Q^{2}$ cuts on the datasets and the data below $Q^2 = Q^2_{\rm min} \leq 6.5$ have been excluded form the QCD fit. Overall, one can conclude from these figures that HT has more impact on theoretical predictions at lower $ Q^2 $ values as expected, and becomes gradually ineffective as $ Q^2 $ increases. Another point should be mentioned is that for smaller values of $ Q^2 $, HT has more impact on points with medium values of $ \beta $, while as $ Q^2 $ increases it affects points with larger values of $ \beta $. For example, one can see from Figs.~\ref{fig:RatioQ2.5} and~\ref{fig:RatioQ5.09} that the greatest impact of HT on theoretical predictions related to $ Q^2=2.5 $ and 5.09 GeV$^2$ occurs at $ \beta=1.78\times 10^{-2} $ and $ \beta=5.62\times 10^{-2} $, so that the differences between theory and data are decreased by including HT effects in the analysis. However, for data points with $ Q^2=8.8 $ GeV$^2$ which have been shown in Fig.~\ref{fig:RatioQ8.8}, the greatest impact of HT occurs in the last panel where $ \beta $ has larger values than before, namaly, $ \beta=5.62\times 10^{-1} $.   

As a short summary, we confirm the remarkable impact that HT correction would have on the diffractive PDFs, especially on those of gluon density shape. Our study also suggests that the reductions in the uncertainties of extracted quark and gluon distributions are small. From the results presented in this section, we also find that the most significant effect on the total and individual $\chi^2$ will be achieved with the HT corrections and the much smaller kinematic cut on the $Q^2_{\rm min}$ which lead to the inclusion of more data points in the analysis. In the next section, we summarize and discuss our findings in this analysis and our outlook for future studies in which more stringent constraints can be found on diffractive PDFs.

%
\section{Summary and Conclusions} \label{sec:Discussion}
%

The diffractive process in deep inelastic lepton-proton ($\ell p$) scattering offers a remarkably
versatile tool to probe the structure of proton in term of the quark and gluon density functions.
It is well-known that the QCD factorization theory allows one to model the diffractive DIS cross sections in terms of non-perturbative diffractive PDFs and hard scattering partonic cross sections in such a way that the precise measurements of cross section can be used to extract the diffractive PDFs. The precision on such measurements could impose very stringent constraints on the quark and gluon distributions.
To this end and in order to present a reliable and consistent determination of diffractive PDFs, we analyzed all available diffractive DIS datasets including the recent H1/ZEUS combined diffractive cross section measurements and determined the diffractive PDFs along with their uncertainty bands evaluated within the ``Hessian'' approach.

Our analysis is also enriched with the higher twist (HT) contributions to the power corrections in diffractive DIS which extend to small values of $Q^{2}$. It has been clearly demonstrated in this analysis that the HT terms in diffractive DIS are required for the kinematic coverage of the diffractive DIS processes at HERA. The most noticeable features that emerge from this study are the better fit quality and slightly significant reduction in the $\chi^2$ after considering the HT corrections.

Recent QCD studies clearly show that the key advantage of diffractive DIS data lies in the fact that it provides the possibility of determination of the diffractive gluon and quark density functions. However, the combination of partonic flavors and more constrain on the gluon density also are challenging.
Consequently, the unprecedented precision and kinematic coverage of the diffractive dijet measurements at HERA~\cite{Chekanov:2009ac,Andreev:2015cwa,Andreev:2014yra} and diffractive hadronic processes at the LHC  
certainly enhance our knowledge on diffractive PDFs, and hence, provide new insights into the inner structure of the nucleon in term of its most basic constituents~\cite{Britzger:2018zvv,Rasmussen:2018dgo,Goncalves:2019agu,Helenius:2019gbd}. Another possible area of future research would be to investigate the effect arising from any available diffractive observable and possible theory developments in QCD analysis of diffractive PDFs. Considering any new improvements, future studies on the current topic are therefore recommended.


%
\section*{Acknowledgments}
%

Hossein Mehraban and Atefeh Maktoubian are grateful the Semnan University for financial support of this project.
Hamzeh Khanpour and Muhammad Goharipour thank School of Particles and Accelerators, Institute for Research in Fundamental Sciences (IPM) for financial support of this research. Hamzeh Khanpour also is thankful the University of Science and Technology of Mazandaran for financial support provided for this project, and CERN theory department for their hospitality and support during the preparation of this paper.


%
%

\clearpage

%

\end{document}